\documentclass[aps,pre,preprint,groupedaddress,showpacs,eqsecnum]{revtex4}
\usepackage{graphicx,epsf,dsfont,amssymb,amsmath,verbatim}

\newcommand{\vk}{{\vec{k}}}
\newcommand{\vp}{{\vec{p}}}

\begin{document}

\title{CONSISTENT DESCRIPTION OF KINETICS AND HYDRODYNAMICS OF DUSTY PLASMA}

\author{B.~Markiv}%

\affiliation{Institute for Condensed
Matter Physics of the National Academy of Sciences of Ukraine,
1~Svientsitskii Str., 79011 Lviv, Ukraine}
\author{M. Tokarchuk}
\affiliation{Institute for Condensed Matter Physics of the
National Academy of Sciences of Ukraine,
1~Svientsitskii Str.,  79011 Lviv, Ukraine}
\affiliation{National University ``Lviv Polytechnic'', 12 Bandera Str., 79013 Lviv, Ukraine}



\begin{abstract}
A consistent statistical description of kinetics and hydrodynamics of dusty plasma is proposed based on the Zubarev nonequilibrium statistical operator method. For the case of partial dynamics the nonequilibrium statistical operator and the generalized transport equations for a consistent description of kinetics of
dust particles and hydrodynamics of electrons, ions and neutral
atoms are obtained. In the approximation of weakly nonequilibrium process a spectrum
of collective excitations of dusty plasma is investigated
in the hydrodynamic limit.
\end{abstract}

\maketitle

\section{Introduction}

The study of nonequilibrium properties of dusty plasma is relevant in the fields of controlled thermonuclear fusion, nuclear reactors, plasma-dust structures~\cite{1,2,2a}, low-temperature plasma~\cite{3,4} etc. The difficulties in describing such systems with inherent processes of  self-organization and structuring are related with a large asymmetry in size, mass and charge of components (electrons, ions, neutral atoms and dust particles), effect of neutral particles ionization as well as with adsorption of neutral particles onto the surface of dust particles. In review~\cite{5}
an important analysis of specific properties of dusty plasma including changes in charge of dust particles under the influence of electrons and ions flow is given in comparison with the classical (electron-ion) plasma. Herewith, some features of dusty plasma are manifested. For example, the one is that the own energy of dust particles can be changed by plasma flows and another is that in the potential electric field, force acting on dust particles is not potential, therefore, vortex dust motions can be excited.
The charge of dust particles is not fixed and depends on the plasma flows on their surface. Herewith, convergence or removal of dust particles leads to the changes in plasma flows. This, in turn, leads to a consistent change in charge of dust particles and to a possible change of sign of interaction energy, in particular, to attraction between negatively charged dust grains \cite{5,5a} what, in turn, leads to the formation of dust plasma crystals \cite{2,3,5, Bonitz1,Bonitz2}
This is provided by the plasma flows which serve as a mechanism of accumulation of positive charge of ions between the negatively charged dust particles.

Complex electromagnetic processes of charging and recharging of dust grains,
which entail the appearance of dust particles with a large electric charge, as well as electron-ionic transport processes demand the development of the theory of effective screening potentials of interaction between particles~\cite{6,7}. In such a system kinetic and hydrodynamic electromagnetic processes should be described consistently.
Keeping this in mind the kinetic theory of dusty plasma was developed be various authors~\cite{44,55,66,8,9,10,11,12,13,33,43,53,63,73}.

In particular, a statistical theory of dusty plasma was formulated in Ref.~\cite{66}, where
the microscopic equations for phase densities and BBGKY hierarchy for the nonequilibrium distribution functions of electrons, ions, neutral atoms and  dust particles were obtained.
On this basis  the kinetic theory of electromagnetic fluctuations in dusty plasma was developed in Ref.~\cite{14} and it was shown that such a description leads to the dependence of the effective cross section of dust charging in wave vector. This reflects the influence of plasma inhomogeneity effects. One of the main problems consist in  the investigation of charge fluctuations in dusty plasma. Important results in this field were obtained in Refs.~\cite{15,16,17}. In particular, in~\cite{17} charge fluctuations in dusty plasma were studied by means of the Brownian dynamics computer simulations within the drift-diffusion approximation. Influence of emission on charge and effective potential of dust grain  in plasma at different intensities of external source of ionization were investigated using numerical methods~\cite{18} as well.
The kinetics of atom-ion scattering processes in dusty plasma was studied in a recent paper~\cite{sem} based on the model collision integrals.

A number of papers by Tsitovich and De Angelis~\cite{13,33,43,53,63,73} were devoted to kinetic description of dusty plasma by means of the Bogolubov-Klimontovich approach.
Here, dusty plasma is considered as a system of interacting ions, electrons, dust particles in the surrounding of neutral component (atoms), which is not taken into account explicitly. A special feature of such an approach consist in considering plasma flows onto dust particles using the charge of dust particles as additional variable \cite{13}.
Thus, $ \frac{d}{dt}q=I_{ext}+\sum_{a}I_{a}$,   where $\sum_{a}I_{a}$ are the flows of plasma particles onto the surface of grains and $I_{ext}$  includes flows of photoelectron and secondary electron emission from their surface. The investigations of electrons and ions distribution in dusty plasma is actively carried out
for the description of long-range interactions and memory effects within the nonclassical
theory~\cite{Trib1,Trib2,Du} based on Tsallis and Renyi entropies.

The study of time correlation functions and transport coefficients, in particular, viscosity, heat conductivity, thermodiffusion, as well as ionic and electronic conductivity
of dusty plasma being an open spatially inhomogeneous system \cite{Fort1} is another important problem. Since dusty plasma is characterized by large asymmetry in size, charge and mass of the components, obviously, the dynamics of each component in transport processes is corresponding. That is to be expected that the time correlation functions ``density-density'', ``momentum-momentum'' as well as transport coefficients of viscosity and heat conductivity for each component, ionic and electron conductivity have  its inherent time and spatial behavior. This behavior latter is affected by kinetic processes related to the dynamics of dust particles.

In the present work, we propose a consistent description of kinetics and hydrodynamics of dusty plasma by means of the Zubarev nonequilibrium statistical operator (NSO) method~\cite{19,19'}. We consider description of the system in terms of both partial and conservative dynamic variables without taking into account processes of ionization and
atoms adsorption onto grains surface. In the second section the nonequilibrium statistical operator is built and based on it the set of transport equations for a consistent description of kinetics of dust particles and hydrodynamics of electrons, ions and neutral atoms was obtained. In Sec.~III we consider the case of weakly nonequilibrium processes, there the system of equations for time correlation functions was received and analyzed. In Sec.~IV by means of the perturbation theory for collective excitations~\cite{23,25} a spectrum of collective modes in the system is investigated in the hydrodynamic limit. The obtained results are shortly concluded in Sec~V.

\section{Nonequilibrium statistical operator and transport equations}

Let us consider dusty plasma as a system of $N_e$ electrons, $N_a$ atoms of species  $a$, $N_i$ ions of species $i$ and $N_d$ dust particles that interact. Hamiltonian of such a multicomponent system can be presented in the form:
\begin{eqnarray}\label{math/IV.1}
H(t)=H_{e}+H_i+H_{a}+H_{d} +\sum_{\alpha=e,i,d}\sum_{s=1}^{N_{\alpha}}Z_{s}e \varphi ({\vec r}_{s};t) ,
\end{eqnarray}
where
\begin{eqnarray}\label{math/IV.2}
H_{\alpha}=\sum_{s=1}^{N_{\alpha}}\frac{(p_s^\alpha)^2}{2m_{\alpha}}+\sum_{s=1}^{N_{d}}\frac{Z^{2}_{s}e^{2}}{a_{s}}
+\frac{1}{2}\sum_{\gamma}\sum_{s\ne
s'}^{N_{\alpha},N_{\gamma}}V_{{\alpha}{\gamma}}(|{\vec r}_{ss'}|)
\end{eqnarray}
is the Hamiltonian of electron $(e)$, ionic $(i)$, atom $(a)$ and dusty $(d)$ subsystems, respectively. $\frac{Z^{2}_{s}e^{2}}{a_{s}}$ is the internal energy of grains,
$Z_{s=d}$ is their valency and $a_{s=d}$ is their size;
\[
{\vec p}_s^\alpha={\vec
p}_s^{\prime\alpha}-\delta_\alpha\frac{Z_\alpha e}{c}{\vec
A}({\vec r};t),
\]
\[
\delta_\alpha=\left\{%
\begin{array}{ll}
    1, & \hbox{{$\alpha=i,e,d$};} \\
    0, & \hbox{{$\alpha=a$}.} \\
\end{array}%
\right.
\]
Here, ${\vec p}_\alpha$ is the impulse of particle ${\alpha=e,i,a,d}$, ${\vec A}({\vec r};t)$ is the vector potential and $\varphi ({\vec r};t)$ is the scalar potential of electromagnetic field induced by charged particles. Expressions for  potentials of interaction $V_{\alpha\alpha}$ as well as for interspecies energy of interaction  $V_{ei}$,
$V_{ea}$, $V_{ed}$, $V_{ia}$, $V_{id}$, $V_{ad}$ are presented in \cite{3}.

The nonequilibrium state of such a multicomponent system of charged and neutral particles
is related with the dynamics of each component as well as with interaction between them.
Different components can remain in kinetic or hydrodynamic, stationary or nonstationary state. Herewith, we can say about quasineutrality of a dusty plasma.
Let us consider the nonequilibrium state of the system when subsystem of electrons, ions and neutral atoms is in hydrodynamic state and dust particles whose mass and charge can change remain in kinetic state.
The parameters of dusty plasma can be the following:
densities of ions and electrons
$n_{i}\approx n_{e}\approx 10^{8}\div 10^{10}$~cm$^{-3}$,
density of atoms $n_{a}\approx 10^{12}\div 10^{17}$~cm$^{-3}$,
density of dust particles
$n_{d}\approx 10^{3}\div 10^{5}$~cm$^{-3}$,
temperature of electrons $T_{e}\approx 1\div 3 \text{еВ}$  and ions $T_i/T_e=10^{-2}$, as well as the charge  $Z_d \approx 3 \cdot 10^3 \div 10^4$ and size $a\approx 5\div 10$~$\mu$m of dust particles.
Since dusty plasma is composed of particles with large asymmetry in masses and charges the characteristic times of appropriate subsystems will differ significantly.
It is important to note the features of dusty plasma related to the fact that large charge of dust particles (which is not constant and depends on plasma flows on the grain surface) lead to the emergence of collective plasma flows, influence of which is comparable or even larger than influence of electrostatic field \cite{5}.
Beside this, the internal energy of grains $\frac{Z_{d}^{2}e^{2}}{2a_{d}}$ exceeds much their kinetic energy and average energy of their interaction. In the potential electric field, forces acting on dust particles are not potential: $[\vec{\nabla}\times Z_{d}\vec{E}]=[(\vec{\nabla}Z_{d})\times \vec{E}]\neq 0$ and vortex motions are observed in the majority of experiments. The interaction of dust particles can be strong enough in distances much more than screening length, and the electrostatic potential of grain absorbing a plasma flow decays as $\frac {1}{r^{2}}$ at large distance. At certain conditions this provides the emergence of dust-plasma crystals and other dissipative self-organized structures.

That is why as a reduced-description parameters of such a nonequilibrium state it is suitable to chose partial variables: averaged values of densities of particles number, momentum and energy of electrons, ions and atoms for the description of hydrodynamic state
\begin{eqnarray}\label{math/IV.3}
\langle{\hat{n}_\alpha({\vec r})}\rangle^t, \qquad
\langle{\hat{\vec \jmath}_\alpha({\vec r})}\rangle^t, \qquad
\langle{\hat{\varepsilon}_\alpha({\vec r})}\rangle^t.
\end{eqnarray}
Here, we introduce the notation
$\langle...\rangle^t=\int{}d\Gamma_N...\varrho(t)$.
\begin{eqnarray}\label{math/IV.4}
\hat{n}_\alpha({\vec r})=\sum_{s=1}^{N_\alpha}\delta ({\vec
r}-{\vec r}_s)
\end{eqnarray}
is the microscopic number density of particles of species ${\alpha=\{e,i,a\}}$;
\begin{eqnarray}\label{math/IV.5}
\hat{\vec \jmath}_\alpha({\vec r})=\sum_{s=1}^{N_\alpha}{\vec
p}_s\delta ({\vec r}-{\vec r}_s)
\end{eqnarray}
is the microscopic momentum  density of particles of species $\alpha$;
\begin{eqnarray}\label{math/IV.6}
\hat{\varepsilon}_\alpha({\vec
r})=\sum_{s=1}^{N_\alpha}\left(\frac{p_s^2}{2m_\alpha} +\frac{Z^{2}_{s}e^{2}}{a_{s}} +\frac{1}{2}\sum_{s'\ne
s=1}^{N_\alpha}\Phi_{\alpha\alpha}(|{\vec r}_s-{\vec
r}_{s'}|)\right)\delta ({\vec r}-{\vec r}_s)
\end{eqnarray}
is the microscopic energy density of particles of species $\alpha$.

For the description of kinetics of dust particles we can chose the nonequilibrium distribution function
\begin{eqnarray}\label{math/IV.7}
 \langle{\hat{n}_d({\vec r},{\vec
p},Z)}\rangle^t,
\end{eqnarray}
where
\begin{eqnarray}\label{math/IV.8}
\hat{n}_d({\vec r},{\vec p},Z)=\sum_{s=1}^{N_d}\delta ({\vec
r}-{\vec r}_s )\delta ({\vec p}-m_s{\vec v}_s)\delta (Z-Z_s)
\end{eqnarray}
is the microscopic phase density of charged grains, which takes into account changes
in mass and charge of particle in the charging/recharging processes.
The set of partial dynamic variables allow us to study processes of  kinetic origin,
for example, relaxation processes related to the difference of partial temperatures of subsystems. This is especially important beyond the hydrodynamic regime.
Such relaxation processes cannot be investigated using the set of conservative variables.

Motions of electrons, ions and charged grains induce corresponding electromagnetic fields, which, in turn,  cause  processes of polarization and change dielectric properties of whole the system. Thus, besides (\ref{math/IV.3}), (\ref{math/IV.7}), as an additional parameters of the reduced description should be the averaged value of electric and magnetic fields $\langle{\hat{\vec E}({\vec r})}\rangle^t$, $\langle{\hat{\vec B}({\vec r})}\rangle^t$ induced by electrons, ions and charged dust particles as well as the their inductions $\langle{\hat{\vec D}({\vec r})}\rangle^t$, $\langle{\hat{\vec
H}({\vec r})}\rangle^t$, which satisfy the averaged Maxwell equations:
%
%
\begin{eqnarray}\label{math/IV.9}
\vec{\nabla}\cdot\langle\hat{\vec B}({\vec r})\rangle^t=0,
\end{eqnarray}
\begin{eqnarray}\label{math/IV.10}
\vec{\nabla}\cdot\langle\hat{\vec D}({\vec
r})\rangle^t=\sum_{i}\langle{\hat{n}_i({\vec
r})}\rangle^t{}Z_ie+e\langle{\hat{n}_e({\vec
r})}\rangle^t+\int\!\!dZ\!\!\int\!\!d{\vec
p}(Z e)\langle{\hat{n}_d({\vec r},{\vec p},Z_d)}\rangle^t,
\end{eqnarray}
\begin{eqnarray}\label{math/IV.11}
\vec{\nabla}\times\langle\hat{\vec E}({\vec
r})\rangle^t+\frac{\partial}{\partial t}\langle\hat{\vec B}({\vec
r})\rangle^t=0,
\end{eqnarray}
\begin{eqnarray}\label{math/IV.12}\lefteqn{
\vec{\nabla}\times\langle\hat{\vec H}({\vec
r})\rangle^t-\frac{\partial}{\partial t}\langle\hat{\vec D}({\vec
r})\rangle^t=\frac{e}{m_e}\langle{\hat{\vec \jmath}_e({\vec
r})}\rangle^t}
\\&&\mbox{}+\sum_i\frac{Z_ie}{m_i}\langle{\hat{\vec \jmath}_i({\vec
r})}\rangle^t+\int\!\!dZ\!\!\int\!\!d{\vec p}\frac{{\vec
p}}{m_d}(Z e)\langle{\hat{n}_d({\vec r},{\vec
p},Z_d)}\rangle^t.\nonumber
\end{eqnarray}
The electric end magnetic fields connected with the scalar and the vector potentials induced by the charged particles:
\begin{eqnarray}\label{math/IV.111}
\langle\hat{\vec E}({\vec
r})\rangle^t =-\vec{\nabla}\cdot\varphi ({\vec r};t)- \frac{\partial}{\partial t} {\vec A}({\vec r};t),
\end{eqnarray}
\begin{eqnarray}\label{math/IV.111}
\langle\hat{\vec B}({\vec r})\rangle^t=\vec{\nabla}\times{\vec A}({\vec r};t).
\end{eqnarray}
As we can see from the structure of equations (\ref{math/IV.9})--(\ref{math/IV.12}), the parameters of reduced description (\ref{math/IV.3}), (\ref{math/IV.7}) are not independent of $\langle{\hat{\vec E}({\vec r})}\rangle^t$, $\langle{\hat{\vec
B}({\vec r})}\rangle^t$, $\langle{\hat{\vec D}({\vec r})}\rangle^t$, $\langle{\hat{\vec H}({\vec r})}\rangle^t$. Contrary, field-particle transport processes are interconnected
and should be considered consistently. Moreover, coordination of hydrodynamics of electrons, ions and atoms with kinetics of dust particles occurs not only through the generalized transport equations, but also in the formation of electromagnetic field (\ref{math/IV.10}),
(\ref{math/IV.12})~\cite{22}. Beside this, known integral relations between
$\langle{\hat{\vec D}({\vec r})}\rangle^t$, $\langle{\hat{\vec E}({\vec r})}\rangle^t$ and $\langle{\hat{\vec
B}({\vec r})}\rangle^t$, $\langle{\hat{\vec H}({\vec
r})}\rangle^t$ determine spatially inhomogeneous dielectric function
$\varepsilon({\vec r},{\vec r}';t,t')$ and magnetization
$\chi({\vec r},{\vec r}';t,t')$ describing polarization processes in the system

Averaged values of densities of particles number, momentum and energy of electrons, ions, atoms and phase density of charged grains number (\ref{math/IV.3}), (\ref{math/IV.7}) as well as  averaged values of fields (\ref{math/IV.9})--(\ref{math/IV.12})
are calculated using nonequilibrium statistical operator $\varrho(t)$ that satisfies the corresponding Liouville equation.
In order to find $\varrho(t)$ we make use of the Zubarev NSO method~\cite{19}. This method allows us to write down $\varrho(t)$
in a general form:
\begin{eqnarray}\label{math/IV.13}
\varrho(t)=\varrho_q(t)-\int\limits_{-\infty}^te^{\varepsilon(t'-t)}T(t,t')(1-{\cal
P}_q(t')) iL_N\varrho_q(t')dt',
\end{eqnarray}
where $iL_N$ is the Liouville operator corresponding to the Hamiltonian (\ref{math/IV.1});
$T(t,t')=\exp\left\{-\int^{t'}_t(1-{\cal
P}_q(t''))iL_Ndt''\right\}$ is the generalized evolution operator with regard to projection $(1-{\cal P}_q(t''))$; ${\cal P}_q(t'')$ is the generalized Kawasaki-Gunton projection operator, whose structure depends on a form of a quasiequilibrium statistical operator
$\varrho_q(t)$. Within the NSO method, $\varrho_q(t)$ is derived
from the extremum of an informational entropy at the fixed values of the reduced-description parameters (in our case Eqs. (\ref{math/IV.3}),
(\ref{math/IV.7})) including the normalization condition $\int{}d\Gamma\varrho_q(t)=1$:
\begin{align}\label{math/IV.14}
&\varrho_q(t)=\exp\biggl\{-\Phi(t)\\
&-\sum_\alpha\int\!\!d{\vec r}\beta_\alpha({\vec r};t)
\biggl[\hat{\varepsilon}_\alpha({\vec r})-\left({\vec
v}_\alpha({\vec r};t)+\delta_{\alpha}\frac{Z_\alpha e}{m_\alpha
c}{\vec A}({\vec r}; t)\right)\cdot\hat{\vec \jmath}_\alpha ({\vec r}) \nonumber\\
&-\Bigl(\nu_\alpha({\vec r};t)-\frac{m_\alpha v_\alpha^2({\vec
r};t)}{2}\Bigr) \hat{n}_\alpha ({\vec r})\biggr] -\int\!\!d{\vec
r}\!\!\int\!\!d\xi a_d({\vec r},\xi;t)
\hat{n}_d({\vec r},\xi)\biggr\},\nonumber
\end{align}
where a new variable $\xi=({\vec p},Z)$ is introduced, $\Phi(t)$ is the Massieu-Planck functional determined from the normalization condition for $\varrho_q(t)$:
\begin{align}\label{math/IV.15}
&\Phi(t)=\ln\!\int\!\!d\Gamma_N
\exp\biggl\{-\sum_\alpha\int\!\!d{\vec r}\beta_\alpha({\vec
r};t)\\
&\times\biggl[\hat{\varepsilon}_\alpha({\vec r})-\left({\vec
v}_\alpha({\vec r};t)+\delta_{\alpha}\frac{Z_\alpha e}{m_\alpha
c}{\vec A}({\vec r}; t)\right)\cdot\hat{\vec \jmath}_\alpha ({\vec r})
\nonumber\\
& -\Bigl(\nu_\alpha({\vec r};t)-\frac{m_\alpha v_\alpha^2({\vec
r};t)}{2}\Bigr) \hat{n}_\alpha ({\vec r})\biggr]-\int\!\!d{\vec
r}\!\!\int\!\!d\xi a_d({\vec r},\xi;t)\hat{n}_d({\vec r},\xi)\biggr\};\nonumber
\end{align}
$\beta_\alpha({\vec r};t)$ is the inverse local temperature for subsystem $\alpha$; ${\vec v}_\alpha({\vec r};t)$ is the hydrodynamic velocity of component $\alpha$;
\begin{eqnarray}\label{math/IV.16}
\nu_\alpha({\vec r};t)=\mu_\alpha({\vec r};t)+Z_\alpha e\varphi({\vec
r};t);
\end{eqnarray}
$\nu_\alpha({\vec r};t)$ is the electrochemical potential of ions and electrons (at $\alpha=i$ and $\alpha=e$, respectively), $\varphi({\vec
r};t)$ is the electric potential, whose gradient determine a longitudinal component of the electric field induced by ions, electrons and dust grains:
\begin{eqnarray}\label{math/IV.17}
{\vec E}^{l}({\vec r};t)=-\vec\nabla \varphi({\vec r};t)=\langle\hat{\vec
E}({\vec r})\rangle^t;
\end{eqnarray}
$\mu_\alpha({\vec r};t)$ is the local chemical potential of component $\alpha$.
The local thermodynamic parameters
$\{\beta_\alpha({\vec r};t)$, ${\vec v}_\alpha({\vec r};t)$,
$\nu_\alpha({\vec r};t)\}$ are determined from the self-consistency conditions:
\begin{eqnarray}\label{math/IV.18}
\langle\hat{\varepsilon}'_\alpha({\vec r})\rangle^t=
\langle\hat{\varepsilon}'_\alpha({\vec r})\rangle^t_q, \quad
\langle\hat{\vec \jmath}_\alpha({\vec r})\rangle^t=
\langle\hat{\vec \jmath}_\alpha({\vec r})\rangle^t_q, \quad
\langle\hat{n}_\alpha({\vec r})\rangle^t=
\langle\hat{n}_\alpha({\vec r})\rangle^t_q,
\end{eqnarray}
where $\hat{\varepsilon}'_\alpha({\vec
r})=\hat{\varepsilon}_\alpha({\vec r})-\left({\vec v}_\alpha({\vec
r};t)+\delta_{\alpha}\frac{Z_\alpha e}{m_\alpha c}{\vec A}({\vec
r}; t)\right)\cdot\hat{\vec \jmath}_\alpha ({\vec r})+\frac{m_\alpha
V_\alpha^2({\vec r};t)}{2}\hat{n}_\alpha({\vec r})$ is the local value of energy in the reference frame moving together with a system element of the
mass velocity ${\vec v}_\alpha({\vec r};t)$.
These parameters satisfy the corresponding generalized thermodynamic relations.
The parameter $a_d({\vec r},{\vec p};Z;t)$ is conjugated to
$\langle{\hat{n}_d({\vec r},{\vec p},Z)}\rangle^t$ and is determined from the self-consistency condition
\begin{eqnarray}\label{math/IV.19}
\langle\hat{n}_d({\vec r},{\vec p};Z)\rangle^t=
\langle\hat{n}_d({\vec r},{\vec p};Z)\rangle^t_{q} .
\end{eqnarray}
Taking into account a structure of $\varrho_q(t)$ the generalized projection operator ${\cal P}_q(t)$ can be presented in the form:
\begin{eqnarray}\label{math/IV.20}
{\cal P}(t)\rho&=&\biggl(\varrho_q(t)-\sum_{\alpha}\sum_l\int\!\!
d{\vec r}\frac{\delta\varrho_q(t)}{\delta\langle\hat{b}_{\alpha
l}({\vec r})\rangle^t}\\
&-&\int\!\! d{\vec r}\int d{\vec p}\int
dZ\frac{\delta\varrho_q(t)}{\delta\langle\hat{n}_d({\vec r},{\vec
p};Z)\rangle^t}\biggr)\int d\Gamma_N\rho\nonumber\\
&+&\sum_{\alpha}\sum_l\int d{\vec
r}\frac{\delta\varrho_q(t)}{\delta\langle\hat{b}_{\alpha l} ({\vec
r}) \rangle^t}\int d\Gamma_N \hat{b}_{\alpha l}({\vec
r})\rho\nonumber\\
&+&\int d{\vec r}\int d\xi \frac{\delta\varrho_q(t)}{\delta\langle\hat{n}_d({\vec r},\xi)\rangle^t}\int d\Gamma_N\hat{n}_d({\vec r},\xi)\rho,\nonumber
\end{eqnarray}
where l=1,2,3, $\hat{b}_{\alpha 1}=\hat{\varepsilon}_\alpha({\vec
r})$,\ $ \hat{b}_{\alpha 2}=\hat{\vec \jmath}_\alpha({\vec r})$, $
\hat{b}_{\alpha 3}=\hat{n}_\alpha({\vec r})$.
The operator ${\cal P}_q(t)$ possesses the following properties: ${\cal P}_q(t)\varrho(t)=\varrho_q(t)$, ${\cal P}_q(t){\cal P}_q(t')={\cal P}_q(t)$, ${\cal
P}_q(t)\varrho_q(t')=\varrho_q(t)$. Substituting  (\ref{math/IV.14})
into (\ref{math/IV.13}) we obtain the nonequilibrium statistical operator of dusty plasma:
\begin{align}\label{math/IV.21}
&\varrho(t)=\varrho_q(t)+\sum_\alpha\int d{\vec
r}'\int\limits_{-\infty}^te^{\varepsilon(t'-t)}T(t,t')I_{\varepsilon}^\alpha({\vec
r}';t')\beta_\alpha({\vec r}';t')\varrho_q(t')dt'\nonumber\\
&-\sum_\alpha\int\!\!d{\vec
r}'\int\limits_{-\infty}^t\!\!e^{\varepsilon(t'-t)}
T(t,t')I_{\jmath}^\alpha({\vec r}';t')\beta_k({\vec r}';t')\\
&\times\left({\vec v}_\alpha({\vec
r}';t')+\delta_{\alpha}\frac{Z_\alpha e}{m_\alpha c} {\vec
A}({\vec r}'; t')\right)\varrho_q(t')dt'\nonumber\\
&+\int\!\! d{\vec r}'\!\!\int\!\! d\xi'\!\!\!\int\limits_{-\infty}^t\!\!e^{\varepsilon(t'-t)}T(t,t')I_{n}^d({\vec
r}',\xi';t')a_d({\vec r}',\xi';t')\varrho_q(t')dt',\nonumber
\end{align}
where
\begin{eqnarray}\label{math/IV.22}
I_\varepsilon^\alpha({\vec r};t)=(1-{\cal
P}(t))iL_N\hat{\varepsilon}_\alpha({\vec r}) %
\end{eqnarray}
is the generalized flow of energy density;
\begin{eqnarray}\label{math/IV.23}
I_\jmath^\alpha({\vec r};t)=(1-{\cal P}(t))iL_N\hat{\vec \jmath}_\alpha({\vec r})
\end{eqnarray}
is the generalized flow of momentum density;
\begin{eqnarray}\label{math/IV.24}
I_{n}^d({\vec r},\xi;t)=(1-{\cal P}(t))iL_N\hat{n}_d({\vec
r},\xi)
\end{eqnarray}
is the generalized flow of microscopic density of dust particles.

In Eqs. (\ref{math/IV.22})--(\ref{math/IV.24}) the generalized Mori projection operator ${\cal P}(t)$ has the following structure:
\begin{eqnarray}\label{math/IV.25}
{\cal
P}(t)\hat{A}&=&\langle\hat{A}\rangle^t_q+\sum_{\alpha}\sum_l\int\!\!{}
d{\vec
r}\frac{\delta\langle\hat{A}\rangle^t_q}{\delta\langle\hat{b}_{\alpha
l}({\vec r})\rangle^t}(\hat{b}_{\alpha l}\bigl({\vec
r})-\langle\hat{b}_{\alpha l}({\vec r})\rangle^t\bigr)\\
&+&\int\!\!{} d{\vec
r}\int \xi \frac{\delta\langle\hat{A}\rangle^t_q}{\delta\langle\hat{n}_d({\vec
r},\xi)\rangle^t}\bigl(\hat{n}_d({\vec r},\xi)-\langle\hat{n}_d({\vec r},\xi)\rangle^t\bigr).\nonumber
\end{eqnarray}
Herewith, ${\cal P}(t){\cal P}(t')={\cal P}(t)$, ${\cal
P}(t)(1-{\cal P}(t'))=0$, ${\cal P}(t)\hat{b}_{\alpha l}({\vec
r})=\hat{b}_{\alpha l}({\vec r})$, ${\cal P}(t)\hat{n}_d({\vec
r},{\vec p};Z)=\hat{n}_d({\vec r},{\vec p};Z)$.
Using the NSO  $\varrho(t)$ we can obtain the set of transport equations
for the reduced-description parameters. Let us present it in a matrix form:
\begin{eqnarray}\label{math/IV.26}
\frac{\partial}{\partial t}\langle\tilde{A}({\vec
x})\rangle^t=\langle\tilde{A}({\vec x})\rangle^t_q-\int\!\!{}
d{\vec
x}'\int\limits_{-\infty}^t\!\!e^{\varepsilon(t'-t)}\tilde{\varphi}_{AA}({\vec
x},{\vec x}';t,t')\tilde{F}_A({\vec x};t')dt',
\end{eqnarray}
where $\tilde{A}({\vec x})=col(\hat{b}_{\alpha l}({\vec
r}),\hat{n}_d({\vec r},{\vec p};Z))=col(\hat{n}_\alpha({\vec
r}),\hat{{\vec \jmath}}_\alpha({\vec
r}),\hat{\varepsilon}_\alpha({\vec r}),\hat{n}_d({\vec r},{\vec
p};Z)) $ and $\tilde{F}_A({\vec x};t')=col\Bigl(
-\beta_\alpha({\vec r};t)(\nu_\alpha({\vec r};t)-m_\alpha
v_\alpha^2({\vec r};t)/2)$, $-\beta_\alpha({\vec r};t)\left[{\vec
v}_\alpha({\vec r};t)+\delta_{\alpha}\frac{Z_\alpha e}{m_\alpha
c}{\vec A}({\vec r}; t)\right]$, $\beta_\alpha({\vec
r};t),a_d({\vec r},{\vec p};Z;t)\Bigr)$ are the vector-columns;
\begin{eqnarray}\label{math/IV.27}
\tilde{\varphi}_{AA}({\vec x},{\vec
x}';t,t')&=&\langle\tilde{I}({\vec x};t) T(t,t')\tilde{I}^+({\vec
x}';t')\rangle_q^{t'}\\
&=&\left[
\begin{array}{cccc}
\tilde{0}&\tilde{0}&\tilde{0}&\tilde{0}\\
\tilde{0}&\tilde{\varphi}_{_{I_{\jmath}I_{\jmath}}}&\tilde{\varphi}_{_{I_{\jmath}I_\varepsilon}}
&\tilde{\varphi}_{_{I_{\jmath}I_n^d}}\\
\tilde{0}&\tilde{\varphi}_{_{I_\varepsilon{}I_{\jmath}}}
&\tilde{\varphi}_{_{I_\varepsilon{}I_\varepsilon}}&\tilde{\varphi}_{_{I_\varepsilon{}I_n^d}}\\
\tilde{0}&\tilde{\varphi}_{_{I_n^dI_{\jmath}}}&\tilde{\varphi}_{_{I_n^dI_\varepsilon}}&
\tilde{\varphi}_{_{I_n^dI_n^d}}
\end{array}
\right]\nonumber
\end{eqnarray}
is a block matrix within which $\tilde{\varphi}_{_{I_AI_A}}$ are the matrices
of the generalized transport kernels, namely:
\begin{eqnarray}\label{math/IV.28}
\tilde{\varphi}_{_{I_{\jmath}I_{\jmath}}}({\vec r},{\vec
r}',t,t')= \left[\begin{array}{ccc}
\varphi_{_{I_{\jmath}I_{\jmath}}}^{ee}&\varphi_{_{I_{\jmath}I_{\jmath}}}^{ei}&\varphi_{_{I_{\jmath}I_{\jmath}}}^{ea}\\[5pt]
\varphi_{_{I_{\jmath}I_{\jmath}}}^{ie}&\varphi_{_{I_{\jmath}I_{\jmath}}}^{ii}&\varphi_{_{I_{\jmath}I_{\jmath}}}^{ia}\\[5pt]
\varphi_{_{I_{\jmath}I_{\jmath}}}^{ae}&\varphi_{_{I_{\jmath}I_{\jmath}}}^{ai}&\varphi_{_{I_{\jmath}I_{\jmath}}}^{aa}
\end{array}
\right]_{(r,r';t,t')},
\end{eqnarray}
where diagonal elements are the transport kernels determining the generalized
viscosity coefficients of electrons, ions and atoms; nondiagonal elements are the transport kernels describing the dissipative correlations between flows of momentum density. Similarly, in the matrix
\begin{eqnarray}\label{math/IV.29}
\tilde{\varphi}_{_{I_\varepsilon{}I_\varepsilon}}({\vec r},{\vec
r}',t,t')= \left[\begin{array}{ccc}
\varphi_{_{I_\varepsilon{}I_\varepsilon}}^{ee}&\varphi_{_{I_\varepsilon{}I_\varepsilon}}^{ei}
&\varphi_{_{I_\varepsilon{}I_\varepsilon}}^{ea}\\[5pt]
\varphi_{_{I_\varepsilon{}I_\varepsilon}}^{ie}&\varphi_{_{I_\varepsilon{}I_\varepsilon}}^{ii}
&\varphi_{_{I_\varepsilon{}I_\varepsilon}}^{ia}\\[5pt]
\varphi_{_{I_\varepsilon{}I_\varepsilon}}^{ae}&\varphi_{_{I_\varepsilon{}I_\varepsilon}}^{ai}
&\varphi_{_{I_\varepsilon{}I_\varepsilon}}^{aa}\\[5pt]
\end{array}\right]_{(r,r';t,t')}
\end{eqnarray}
the diagonal elements are the transport kernels determining the generalized
heat conductivity coefficients of electrons, ions and atoms subsystems; nondiagonal elements are the transport kernels describing the dissipative correlations between flows of energy density of particles. Correspondingly, the transport kernels of matrices
$\tilde{\varphi}_{_{I_\varepsilon{}I_{\jmath}}}({\vec r},{\vec
r}',t,t')$, $\tilde{\varphi}_{_{I_{\jmath}I_\varepsilon}}({\vec
r},{\vec r}',t,t')$ describe dissipative correlations between the generalized flows of
momentum and energy of electrons, ions and atoms.
The transport kernels of matrices
$\tilde{\varphi}_{_{I_{\jmath}I_n^d}}({\vec r},{\vec r}',{\vec
p}';Z;t,t')$, $\tilde{\varphi}_{_{I_n^dI_{\jmath}}}({\vec r},{\vec
r}',{\vec p}';Z;t,t')$,
$\tilde{\varphi}_{_{I_n^dI_\varepsilon}}({\vec r},{\vec r}',{\vec
p}';Z;t,t')$, $\tilde{\varphi}_{_{I_\varepsilon{}I_n^d}}({\vec
r},{\vec r}',{\vec p}';Z;t,t')$ in Eq. (\ref{math/IV.27}) describe,
respectively, dissipative correlations between the generalized flows of momentum and energy density (of electrons, ions and atoms) and the generalized flow of microscopic phase density of dust particles. Herewith,
$$\tilde{I}({\vec x};t)=col(I^\alpha_n({\vec r};t),I_{\jmath}^\alpha({\vec r};t),
I^\alpha_\varepsilon{}({\vec r};t),I_n^d({\vec r},{\vec p};Z;t))$$
is the column-vector,
$$\tilde{I}^{(+)}({\vec x}';t') = (I^\alpha_n({\vec r}';t'),I_{\jmath}^\alpha({\vec r}';t'),
I^\alpha_\varepsilon{}({\vec r}';t'),I_n^d({\vec r}',{\vec
p}';Z';t'))
$$
is the row-vector, and $\tilde{I}({\vec x}';t')\tilde{I}^{(+)}({\vec
x}';t')$ is their scalar product. The transport kernel
$\tilde{\varphi}_{_{I_n^d I_n^d}}({\vec r},{\vec p};Z;{\vec
r}',{\vec p}';Z';t,t')$ describes dissipative correlations between the generalized flows of microscopic phase density of grains and determines the generalized diffusion coefficient in phase space of grains.

Motions of electrons, ions and charged dust particles, according to Maxwell equations
(\ref{math/IV.9})--(\ref{math/IV.12}), induce electromagnetic fields.
The dissipative transport processes related to the flows of momentum and energy densities of electrons, ions, atoms and kinetics of charged grains that are described by the set of the nonmarkovian transport equations affect on the dissipation of field variables through the right side of equations (\ref{math/IV.10}), (\ref{math/IV.12}).

In our approach, effects of adsorption/desorbtion of electrons, ions and atoms at the grain surface as well as processes of clustering of dust particles can be described by means of the nonequilibrium correlation functions
$\langle \hat{n}_d({\vec r},\xi)\hat{n}_{\alpha}({\vec r}')\rangle^{t}=\langle \hat{G}_{nn}^{d \alpha}({\vec r},\xi;{\vec r}')\rangle^{t}$, $\langle \hat{n}_d({\vec r},\xi)\hat{n}_{d}({\vec r}',\xi')\rangle^{t}=\langle \hat{G}_{nn}^{d d}({\vec r},\xi;{\vec r}',\xi')\rangle^{t}$, where $\hat{G}_{nn}^{d \alpha}({\vec r},\xi;{\vec r}')=\hat{n}_d({\vec r},\xi)\hat{n}_{\alpha}({\vec r}')$ and
$\langle \hat{G}_{nn}^{d d}({\vec r},\xi;{\vec r}',\xi')=\hat{n}_d({\vec r},\xi)\hat{n}_{d}({\vec r}',\xi')$.
Generally speaking, taking into account the fact that electrons are localized on the grain's surface, such a description should be conducted at quantum level.
Then, in the abovementioned correlation functions, $ \hat{n}_d({\vec r},\xi)$ and $\hat{n}_{\alpha}({\vec r}')$
are the density operators of particles of corresponding species. In particular,
$\hat{n}_{a}({\vec r})=\sum_{j\nu}^{N_{a}}\hat{\Psi}_{j\nu}^{+}({\vec r})
\hat{\Psi}_{j\nu}({\vec r})$,
$\hat{n}_{e}({\vec r})=\sum_{j{\vec s}}^{N_{a}}\hat{\Psi}_{j{\vec s}}^{+}({\vec r})
\hat{\Psi}_{j{\vec s}}({\vec r})$, where $\hat{\Psi}_{j\nu}^{+}({\vec r})$ and
$\hat{\Psi}_{j\nu}({\vec r})$ are the operators of creation and annihilation
of atom in the state $\nu$ on the grain's surface, whereas
$\hat{\Psi}_{j{\vec s}}^{+}({\vec r})$ and
$\hat{\Psi}_{j{\vec s}}({\vec r})$ are the operators of creation and annihilation of
electron in state with spin ${\vec s}$.
The ion interacting with the electron from the dust particles surface create the atom, which
can be in the adsorbed state on the surface or can desorb into ion-electron-atom
subsystem until next ionization. From this point of view, a quantum nature of interaction
of ions, electron and atoms with the electron surface of grain should be taken into account
in the Hamiltonian  (\ref{math/IV.1}). Similar problems appear in the theory of catalytic
processes \cite{kost}.
Then, quasiequilibrium statistical operator (\ref{math/IV.14}) will have the following form:
$$
\varrho_q(t)=\exp\biggl\{-\Phi(t)
-\sum_\alpha\int\!\!d{\vec r}\beta_\alpha({\vec r};t)
\biggl[\hat{\varepsilon}_\alpha({\vec r})-\left({\vec
v}_\alpha({\vec r};t)+\delta_{\alpha}\frac{Z_\alpha e}{m_\alpha
c}{\vec A}({\vec r}; t)\right)\cdot\hat{\vec \jmath}_\alpha ({\vec r})
$$
$$
-\Bigl(\nu_\alpha({\vec r};t)-\frac{m_\alpha v_\alpha^2({\vec
r};t)}{2}\Bigr) \hat{n}_\alpha ({\vec r})\biggr] -\int\!\!d{\vec
r}\!\!\int\!\!d\xi a_d({\vec r},\xi;t)
\hat{n}_d({\vec r},\xi)
$$
$$
-\sum_\alpha\int\!\!d{\vec r}\int\!\!d{\vec
r}'\!\!\int\!\!d\xi \mu_{d \alpha}({\vec r},\xi;{\vec r}';t)
\hat{G}_{nn}^{d \alpha}({\vec r},\xi;{\vec r}')
$$
$$
-\sum_\alpha\int\!\!d{\vec r}\int\!\!d{\vec
r}'\!\!\int\!\!d\xi \int\!\!d\xi' \mu_{d d}({\vec r},\xi;{\vec r}',\xi';t)
\hat{G}_{nn}^{d d}({\vec r},\xi;{\vec r}',\xi')
\biggr\}
$$
and the transport equations (\ref{math/IV.26}) will contain the nonequilibrium correlation functions $\langle \hat{G}_{nn}^{d \alpha}({\vec r},\xi;{\vec r}')\rangle^{t}$, $\langle \hat{G}_{nn}^{d d}({\vec r},\xi;{\vec r}',\xi')\rangle^{t}$.
The parameters $\mu_{d \alpha}({\vec r},\xi;{\vec r}';t)$,
$\mu_{d d}({\vec r},\xi;{\vec r}',\xi';t)$ are determined from the corresponding
self-consistency conditions:
$\langle \hat{G}_{nn}^{d \alpha}({\vec r},\xi;{\vec r}')\rangle^{t}
=\langle \hat{G}_{nn}^{d \alpha}({\vec r},\xi;{\vec r}')\rangle^{t}_{q}$ and
$\langle \hat{G}_{nn}^{d d}({\vec r},\xi;{\vec r}',\xi')\rangle^{t}
=\langle \hat{G}_{nn}^{d d}({\vec r},\xi;{\vec r}',\xi')\rangle^{t}_{q}$
and can be defined as spatially-temporal chemical potential of
corresponding ``dimers''. This issue needs a separate study.

Beside the statistical theory of dusty plasma based on the BBGKY hierarchy
for nonequilibrium distribution functions~\cite{6}, we proposed a consistent description
of kinetic and hydrodynamic processes with partial contribution from each component by means of the Zubarev NSO method. Based on the obtained system of transport equations
(\ref{math/IV.26}) and averaged Maxwell equations~\cite{22} the time correlation functions as well as the generalized transport coefficients of dusty plasma can be investigated for both weakly and strongly nonequilibrium processes. Such a set of equations takes into account consistently kinetic and hydrodynamic nonmarkovian processes as well as mutual influence of  dynamics of particles and electromagnetic field.
The investigation of dependencies of the generalized transport coefficients (viscosity, heat conductivity, ionic and electron conductivity) related with transport kernels (\ref{math/IV.28}), (\ref{math/IV.29}) on wave vector and frequency remains very important.
In this direction main problem can be bring to the calculation of generalized transport kernels (\ref{math/IV.27}), in particular
$\varphi_{I_n^dI_n^d}(t)$, in the kinetic equation for
$\langle\hat{n}_d({\vec r},{\vec p},Z_d)\rangle^t$.
At small concentrations of dust particles, in the transport kernel  $\varphi_{I_n^dI_n^d}(t)$
an expansion over the grains density can be used. At the same time,
an important issue consist in the investigation of weakly nonequilibrium processes in the system, when gradients of nonequilibrium thermodynamic parameters  $\tilde{F}_A({\vec x};t')$ are small. In the linear approximation in deviations of the nonequilibrium thermodynamic parameters from their equilibrium values, the set of transport equations (\ref{math/IV.26})
become closed. As it is known, within the framework of the NSO method, the time correlation functions built on the basic set of dynamic variables of reduced description Eqs. (\ref{math/IV.4})--(\ref{math/IV.6}), (\ref{math/IV.8}) satisfy the same system of equations. This issue we consider in the following section.

\section{Weakly nonequilibrium processes in dusty plasma}

Let us now consider the nonequilibrium processes in dusty plasma when the nonequilibrium thermodynamic parameters $\tilde{F}_A(x;t)$ slightly deviate from their equilibrium values $\tilde{F}_A(x;0)$.
This is equivalent the fact that $\langle\tilde{A}(x)\rangle^t$ slightly deviate from their equilibrium values $\langle\tilde{A}(x)\rangle_0$. Here, $\langle\ldots\rangle_0=\int
d\Gamma_N\ldots\varrho_0$, $\varrho_0$ is the equilibrium statistical operator of dusty plasma. Then, expanding the quasiequilibrium statistical operator Eq. (\ref{math/IV.14}) over deviations $\delta\tilde{F}_A(x;t)=\tilde{F}_A(x;t)-\tilde{F}_A(x;0)$
we restrict ourself to the linear approximation
\begin{eqnarray}
\label{math/IV.39} \varrho_q(t)=\varrho_0\left\{1-\int dx
\delta\tilde{F}_A(x;t)\tilde{A}(x)\right\}.
\end{eqnarray}

Further, we use Fourier transformation. Then, excluding parameters
$\delta\tilde{F}_A(\vec{k};t)$ from
Eq. (\ref{math/IV.39}) by means of the self-consistency conditions
Eqs. (\ref{math/IV.18})--(\ref{math/IV.19}), we obtain
\begin{eqnarray}
\label{math/IV.40} \varrho^0_q(t)=\varrho_0\left\{1+\sum_{\vec{k}}
\int dx\int dx'
\delta\tilde{A}_{\vec{k}}(x;t)\tilde{\Phi}^{-1}_{AA}(\vec{k},x,x')
\tilde{A}_{\vec{k}}(x')\right\},
\end{eqnarray}
where $\langle\delta\tilde{A}_{\vec{k}}(x)\rangle^t=
\delta\tilde{A}_{\vec{k}}(x;t)$,
$\delta\tilde{A}_{\vec{k}}(x)=\tilde{A}_{\vec{k}}(x)
-\langle\tilde{A}_{\vec{k}}(x)\rangle_0$. Henceforth
$x=\{\vec{p},Z\}$  and integration concerns only to a variable describing the grains subsystem.
$\tilde{\Phi}^{-1}_{AA}(\vec{k},x,x')$ is the matrix inverse to the matrix of correlation functions of variables $\tilde{A}_{\vec{k}}(x)$:
\begin{eqnarray}
\label{math/IV.41}
\tilde{\Phi}_{AA}(\vec{k},x,x')&=&\langle\tilde{A}_{\vec{k}}(x)
\tilde{A}^{+}_{\vec{k}}(x')\rangle_0 \\
&=&\left[
\begin{array}{cccc}
\tilde{\Phi}_{nn}&\tilde{0}&
\tilde{0}&\tilde{0}\\
\tilde{0}&\tilde{\Phi}_{pp}&\tilde{0}&\tilde{0}\\
\tilde{0}&\tilde{0}& \tilde{\Phi}_{hh}
&\tilde{0}\\
\tilde{0}&\tilde{0}& \tilde{0} &\tilde{\Phi}_{NN}
\end{array} \right]_{(\vec{k},x,x')}.\nonumber
\end{eqnarray}
Here, in particular,
\begin{eqnarray}
\label{math/IV.42} \tilde{\Phi}_{nn}(\vec{k})=\left[
\begin{array}{ccc}
{\Phi}_{nn}^{ee}&{\Phi}_{nn}^{ei}&{\Phi}_{nn}^{ea}\\
{\Phi}_{nn}^{ie}&{\Phi}_{nn}^{ii}&{\Phi}_{nn}^{ia}\\
{\Phi}_{nn}^{ae}&{\Phi}_{nn}^{ai}&{\Phi}_{nn}^{aa}
\end{array} \right]_{(\vec{k})}
\end{eqnarray}
is the matrix of the equilibrium correlation functions ``density-density''
for electrons, ions and neutral atoms
$\Phi_{nn}^{\alpha\gamma}(\vec{k})=\langle\hat{n}^{\alpha}_{\vec{k}}
\hat{n}^{\gamma}_{\vec{k}}\rangle_0$.
\begin{eqnarray}
\label{math/IV.43} \tilde{\Phi}_{\jmath\jmath}(\vec{k})=\left[
\begin{array}{ccc}
{\Phi}_{\jmath\jmath}^{ee}&0&0\\
0&{\Phi}_{\jmath\jmath}^{ii}&0\\
0&0&{\Phi}_{\jmath\jmath}^{aa}
\end{array} \right]_{(\vec{k})}
\end{eqnarray}
is the matrix of the equilibrium correlation functions ``momentum-momentum''
$\Phi_{\jmath\jmath}^{\alpha\gamma}(\vec{k})=\langle\hat{\vec{\jmath}}_{\vec{k}}^{\alpha}
\hat{\vec{\jmath}}^{\gamma}_{\vec{k}}\rangle_0$.
%
\begin{eqnarray}
\label{math/IV.44} \tilde{\Phi}_{hh}(\vec{k})=\left[
\begin{array}{ccc}
{\Phi}_{hh}^{ee}&{\Phi}_{hh}^{ei}
&{\Phi}_{hh}^{ea}\\
{\Phi}_{hh}^{ie}&{\Phi}_{hh}^{ii}
&{\Phi}_{hh}^{ia}\\
{\Phi}_{hh}^{ae}&{\Phi}_{hh}^{ai} &{\Phi}_{hh}^{aa}
\end{array} \right]_{(\vec{k})}
\end{eqnarray}
is the matrix of the equilibrium correlation functions ``enthalpy-enthalpy''
for electrons, ions and neutral atoms
$\Phi_{hh}^{\alpha\gamma}(\vec{k})
=\langle\hat{h}_{\vec{k}}^{\alpha}\hat{h}_{\vec{k}}^{\gamma}\rangle_0$.
\[
\hat{h}_{\vec{k}}^\alpha=\hat{\varepsilon}_{\vec{k}}^{\alpha}-
\sum_{\gamma\gamma'}\langle\hat{\varepsilon}_{\vec{k}}^{\alpha}\hat{n}_{\vec{k}}^{\gamma}\rangle_0
[\Phi_{nn}^{-1}(\vec{k})]^{\gamma\gamma'}\hat{n}_{\vec{k}}^{\gamma'}
\]
is the generalized enthalpy of subsystem $\alpha$.
\begin{eqnarray}
\label{math/IV.45}
\tilde{\Phi}_{NN}(\vec{k},x,x')={\Phi}_{NN}(\vec{k},\vec{p},Z,\vec{p}',Z')
=\langle\hat{N}_{\vec{k}}(\vec{p},Z)\hat{N}_{-\vec{k}}(\vec{p}',Z')\rangle_0
\end{eqnarray}
is the equilibrium correlation function ``phase density - phase density'' for dust particles. Herewith, the inverse function
${\Phi}_{NN}^{-1}(\vec{k},\vec{p},Z,\vec{p}',Z')$
should be defined from the integral relation
\begin{align}
\label{math/IV.45'} \int d\vec{p}''\int
dZ''&\Phi_{NN}(\vec{k},\vec{p},Z,\vec{p}'',Z'')
\Phi_{NN}^{-1}(\vec{k},\vec{p}'',Z'',\vec{p}',Z')=
\\
&=\delta(\vec{p}-\vec{p}')\delta(Z-Z').\nonumber
\end{align}
The function  $\Phi_{NN}^{-1}(\vec{k},\vec{p}'',Z'',\vec{p}',Z')$
is calculated in Appendix~A.
The new kinetic variable $\hat{N}_{\vec{k}}(\vec{p},Z)$ is orthogonal to
the partial hydrodynamic variables, and it appears as a result of excluding the thermodynamic parameters:
\begin{eqnarray}
\label{math/IV.N} \hat{N}_{\vec{k}}(\vec{p},Z)=(1-{\cal P}^H)
\hat{n}_{\vec{k}}^d(\vec{p},Z).
\end{eqnarray}
${\cal P}^H$ is the hydrodynamic part of Mori projection operator
(\ref{math/IV.47}).

In approximation Eq. (\ref{math/IV.40}), the nonequilibrium statistical operator
(\ref{math/IV.21}) has the following form:
\begin{eqnarray}
\label{math/IV.46}
\varrho(t)&=&\varrho_q^0(t)-\!\sum_{\vec{k}}\!\!\int\!\!
dx\!\!\int
\!\!dx' \!\!\int_{\infty}^te^{\varepsilon(t'-t)}\\
&\times& T_0(t,t')(1-{\cal P}_0)iL_N\tilde{A}_{\vec{k}}(x)
\tilde{\Phi}_{AA}^{-1}(\vec{k},x,x')
\delta\tilde{A}_{-\vec{k}}(x,t')\varrho_0dt',\nonumber
\end{eqnarray}
where $T_0(t,t')=\exp\{(t'-t)(1-{\cal P}_0)iL_N\}$, ${\cal P}_0$ is the Mori projection operator for weakly nonequilibrium processes, which acts on dynamic variables as follows:
\begin{eqnarray}
\label{math/IV.47} {\cal
P}_0\tilde{B}=\langle\tilde{B}\rangle_0+\int dx\int
dx'\langle\tilde{B}\tilde{A}_{\vec{k}}^+(x)\rangle_0
\tilde{\Phi}_{AA}^{-1}(\vec{k},x,x')\tilde{A}_{\vec{k}}(x')
\end{eqnarray}
It possesses the properties ${\cal P}_0(1-{\cal P}_0)=0$, ${\cal
P}_0\tilde{A}_{\vec{k}}(x)=\tilde{A}_{\vec{k}}(x)$.
As we can see, the NSO is a functional of the averaged values $\langle\tilde{A}_{\vec{k}}(x)\rangle^t$ and the generalized flows $(1-{\cal P}_0)iL_N\tilde{A}_{\vec{k}}(x)$. Using $\varrho(t)$ Eq. (\ref{math/IV.46}) we can obtain
the set of equations for
$\langle\tilde{A}_{\vec{k}}(x)\rangle^t$~\cite{19'}. We present it in
a matrix form:
\begin{eqnarray}
\label{math/IV.48} \lefteqn{\frac{\partial}{\partial
t}\langle\delta\tilde{A}_{\vec{k}}(x)\rangle^t-\int
i\tilde{\Omega}_{AA}(\vec{k},x,x')\langle\delta\tilde{A}_{\vec{k}}(x')\rangle^td
x'}
\\&&\mbox{} +\int
dx'\int_{-\infty}^te^{\varepsilon(t'-t)}\tilde{\varphi}_{AA}^{0}
(\vec{k},x,x';t,t')
\langle\delta\tilde{A}_{\vec{k}}(x')\rangle^{t'}dt'=0.\nonumber
\end{eqnarray}
Here,
\begin{eqnarray}
\label{math/IV.49}
\lefteqn{i\tilde{\Omega}_{AA}(\vec{k},x,x')=\int dx''
\langle\dot{\tilde{A}}_{\vec{k}}(x)\tilde{A}_{\vec{k}}^+(x'')
\rangle_0\tilde{\Phi}_{AA}^{-1}(\vec{k}, x'',x')=}
\\&&\quad\quad\quad\quad\mbox{}=\left[
\begin{array}{cccc}
\tilde{0}&i\tilde{\Omega}_{n\jmath}&\tilde{0}&\tilde{0}\\
i\tilde{\Omega}_{\jmath n}&\tilde{0}&i\tilde{\Omega}_{\jmath h}
&i\tilde{\Omega}_{\jmath N}\\
\tilde{0}&i\tilde{\Omega}_{h\jmath}&\tilde{0}&\tilde{0}\\
\tilde{0}&i\tilde{\Omega}_{N\jmath}&\tilde{0}& i\tilde{\Omega}_{NN}
\end{array}
\right]_{(\vec{k},x,x')}\nonumber
\end{eqnarray}
is the frequency matrix, whose elements describe the static correlations between densities of particles number, momentum end energy of each component of dusty plasma.
\begin{eqnarray}
\label{math/IV.50}
\lefteqn{\tilde{\varphi}_{AA}^0(\vec{k},x,x';t,t') =\left[
\begin{array}{cccc}
\tilde{0}&\tilde{0}&\tilde{0}&\tilde{0}\\
\tilde{0}&\tilde{\varphi}_{\jmath\jmath}^0&\tilde{\varphi}_{\jmath h}^0&\tilde{\varphi}_{\jmath N}^0\\
\tilde{0}&\tilde{\varphi}_{h \jmath}^0&\tilde{\varphi}_{hh}^0
&\tilde{\varphi}_{hN}^0\\
\tilde{0}&\tilde{\varphi}_{N\jmath}^0&\tilde{\varphi}_{Nh}^0&\tilde{\varphi}_{NN}^0
\end{array}
\right]_{(\vec{k},x,x';t,t')}}
\\&&\mbox{}
=\int dx'' \langle(1-{\cal
P}_0)\dot{\tilde{A}}_{\vec{k}}(x)T_0(t,t')(1-{\cal P}_0)
\dot{\tilde{A}}_{\vec{k}}^+(x'')\rangle_0
\tilde{\Phi}_{AA}^{-1}(\vec{k},x'',x') \nonumber
\end{eqnarray}
is the matrix, whose elements are the transport kernels (memory functions),
describing dissipative processes in dusty plasma, namely, diffusivity, viscosity and heat conductivity. Herewith,
\begin{eqnarray}
\label{math/IV.51} \tilde{\varphi}_{\jmath\jmath}^0(\vec{k};t,t')
=\left[
\begin{array}{ccc}
{\varphi}_{I_{\jmath}I_{\jmath}}^{ee}&{\varphi}_{I_{\jmath}I_{\jmath}}^{ei}&{\varphi}_{I_{\jmath}I_{\jmath}}^{ea}\\
{\varphi}_{I_{\jmath}I_{\jmath}}^{ie}&{\varphi}_{I_{\jmath}I_{\jmath}}^{ii}&{\varphi}_{I_{\jmath}I_{\jmath}}^{ia}\\
{\varphi}_{I_{\jmath}I_{\jmath}}^{ae}&{\varphi}_{I_{\jmath}I_{\jmath}}^{ai}&{\varphi}_{I_{\jmath}I_{\jmath}}^{aa}
\end{array}
\right]_{(\vec{k};t,t')}
\end{eqnarray}
is the matrix with elements describing viscous processes, in particular,
${\varphi}_{I_{\jmath}I_{\jmath}}^{ee}$,
${\varphi}_{I_{\jmath}I_{\jmath}}^{ii}$ and
${\varphi}_{I_{\jmath}I_{\jmath}}^{aa}$ define the generalized viscosity coefficients of electron, ionic and atom component. The nondiagonal elements describe the intercomponent  viscous processes.
\begin{eqnarray}
\label{math/IV.52} \tilde{\varphi}_{hh}^0(\vec{k};t,t') =\left[
\begin{array}{ccc}
{\varphi}_{I_h I_h}^{ee}&{\varphi}_{I_h I_h}^{ei}
&{\varphi}_{I_h I_h}^{ea}\\
{\varphi}_{I_h I_h}^{ie}&{\varphi}_{I_h I_h}^{ii}
&{\varphi}_{I_h I_h}^{ia}\\
{\varphi}_{I_h I_h}^{ae}&{\varphi}_{I_h I_h}^{ai}&{\varphi}_{I_h
I_h}^{aa}
\end{array}
\right]_{(\vec{k};t,t')}
\end{eqnarray}
is the matrix with elements describing heat conductivity processes, namely,
${\varphi}_{I_h I_h}^{ee}$, ${\varphi}_{I_h I_h}^{ii}$ and
${\varphi}_{I_h I_h}^{aa}$ define the generalized heat conductivity coefficients of
electron, ionic and atom subsystems, and the nondiagonal elements describe the intercomponent  heat processes.

It can be shown, that within the NSO method~\cite{19'}
the time correlation functions
\begin{eqnarray}
\label{math/IV.53} \tilde{\Phi}_{AA}(\vec{k},x,x';t)=\langle\tilde{A}_{\vec{k}}(x;t)\tilde{A}^+_{\vec{k}}(x')\rangle_0=\left[
\begin{array}{cccc}
\tilde{\Phi}_{nn}&\tilde{\Phi}_{n\jmath}
&\tilde{\Phi}_{nh}&\tilde{\Phi}_{nN}\\
\tilde{\Phi}_{\jmath n}&\tilde{\Phi}_{\jmath\jmath}
&\tilde{\Phi}_{\jmath h}&\tilde{\Phi}_{\jmath N}\\
\tilde{\Phi}_{h n}&\tilde{\Phi}_{h \jmath}
&\tilde{\Phi}_{hh}&\tilde{\Phi}_{h
N}\\\tilde{\Phi}_{Nn}&\tilde{\Phi}_{N\jmath}
&\tilde{\Phi}_{Nh}&\tilde{\Phi}_{NN}\\
\end{array}
\right]_{(\vec{k},x,x';t)}
\end{eqnarray}
will also satisfy the equations (\ref{math/IV.48}):
\begin{eqnarray}
\label{math/IV.54} \lefteqn{\frac{\partial}{\partial
t}\tilde{\Phi}_{AA}(\vec{k},x,x';t)-\int dx''
i\tilde{\Omega}_{AA}(\vec{k},x,x'')\tilde{\Phi}_{AA}(\vec{k},x'',x';t)}
\\&&\mbox{} +\int
dx''\int_{-\infty}^te^{\varepsilon(t'-t)}
\tilde{\varphi}_{AA}^{0}(\vec{k},x,x'';t,t')
\tilde{\Phi}_{AA}(\vec{k},x'',x';t')dt'=0.\nonumber
\end{eqnarray}
In particular, the matrix of the time correlation functions ``density-density''
$\tilde{\Phi}_{nn}(\vec{k};t)$ is connected with the matrix of partial dynamic structure factors $\tilde{S}(\vec{k};\omega)$
\begin{eqnarray}
\label{math/IV.55}
\tilde{S}(\vec{k};\omega)=\frac{1}{2\pi}\int_{-\infty}^{\infty}e^{i\omega
t}\tilde{\Phi}_{nn}(\vec{k};t) =\left[
\begin{array}{ccc}
S^{ee}&S^{ei}&S^{ea}\\
S^{ie}&S^{ii}&S^{ia}\\
S^{ae}&S^{ai}&S^{aa}\\
\end{array}\right]_{(\vec{k};\omega)}.
\end{eqnarray}
Applying the Laplace transformation to the matrix equation (\ref{math/IV.54}) we obtain
\begin{eqnarray}
\label{math/IV.56}
s\tilde{\Phi}_{AA}(\vec{k};s)
-i\tilde{\Omega}_{AA}(\vec{k})\tilde{\Phi}_{AA}(\vec{k};s)
+\tilde{\varphi}^0_{AA}\tilde{\Phi}_{AA}(\vec{k};s)=-\tilde{\Phi}_{AA}(\vec{k};0),
\end{eqnarray}
where $s=\omega+i\varepsilon$.

It is worth noting that for the variable
$A_{\vec{k}}(x)=\hat{N}_{\vec{k}}(\vec{p},Z)$ in the set of transport equations
an integration over $\vec{p}$ and $Z$ is present
\begin{eqnarray}
\label{math/IV.57}
\lefteqn{s\tilde{\Phi}_{NN}(\vec{k},\vec{p},Z,\vec{p}',Z';s) }
\\&&\mbox{}-\int
d\vec{p}''\int
dZ''i\tilde{\Omega}_{NN}(\vec{k},\vec{p},Z,\vec{p}'',Z'')
\tilde{\Phi}_{NN}(\vec{k},\vec{p}'',Z'',\vec{p}',Z';s)\nonumber
\\&&\mbox{}-i\tilde{\Omega}_{NA}(\vec{k},\vec{p},Z)
\tilde{\Phi}_{AN}(\vec{k},\vec{p}',Z';s)\nonumber
\\&&\mbox{} +\int d\vec{p}''\int
dZ''\tilde{\varphi}^0_{NN}(\vec{k},\vec{p},Z,\vec{p}'',Z'';s)
\tilde{\Phi}_{NN}(\vec{k},\vec{p}'',Z'',\vec{p}',Z';s) \nonumber
\\&&\mbox{}+\tilde{\varphi}^0_{NA}(\vec{k},\vec{p},Z;s
)\tilde{\Phi}_{AN}(\vec{k},\vec{p}',Z';s)=
-\tilde{\Phi}_{NN}(\vec{k},\vec{p},Z,\vec{p}',Z';s ).\nonumber
\end{eqnarray}

The relation ${\rm
det}|zI-i\tilde{\Omega}_{AA}(\vec{k};s)+\tilde{\varphi}_{AA}^0(\vec{k};s)|=0$
determines the collective excitations in dusty plasma caused by electron, ionic and atom interactions. Each component has its own role in sound spreading, in processes of electroconductivity, heat conductivity, and polarization processes.
It manifests in processes of appearing/disappearing of ordered structures,
in polarization processes in dusty plasma through the partial dynamics structure factors
Eq. (\ref{math/IV.55}) and the ``charge-charge'' structure factor for grains $S_{NN}(\vec{k},Z,Z';\omega)=\int d\vec{p}\int
d\vec{p}'S_{NN}(\vec{k},\vec{p},Z,\vec{p}',Z';\omega)$.

\section{Collective excitations}


A considerable asymmetry of dusty plasma in charges and masses makes its description natural in terms of partial dynamics. However, when the concentration of grains is small and its dynamic can be considered at diffusion level, and for the case of isothermal plasma when temperature of all components is close, we can describe  dusty plasma based on the conservative variables.  In particular, we can include into the set of the reduced-description parameters the averaged values of Fourier components of partial number densities, total momentum and total enthalpy. The dust particles subsystem we  will again described by a kinetic variable. The choice of such variables is caused by the fact that further we are interested in collective excitations of the system in the hydrodynamic limit
and such a description is common for description of multicomponent liquids.
Thus,
\begin{eqnarray}\label{math/IV.58}
\hat{n}_{\vec{k}}^{\alpha}=\sum_{l=1}^{N_\alpha}e^{-ikr_l^{\alpha}}, \qquad%
\hat{\vec{\jmath}}_{\vec{k}}=\sum_\alpha
\hat{\vec{\jmath}}_{\vec{k}}^{\
\alpha}=\sum_\alpha\sum_{l=1}^{N_\alpha}p_l^{\alpha}e^{-ikr_l^{\alpha}}
\end{eqnarray}
are the Fourier components of particles number density of species $\alpha$ and total momentum density, respectively,
\begin{eqnarray}\label{math/IV.59}
\hat{h}_{\vec{k}}=\hat{\varepsilon}_{\vec{k}}-\sum_{\alpha\gamma}
\langle\hat{\varepsilon}_{\vec{k}}\hat{n}_{\vec{k}}^{\alpha}\rangle_0
[\Phi_{nn}^{-1}(\vec{k})]^{\alpha\gamma}\hat{n}_{\vec{k}}^{\gamma}
\end{eqnarray}
are the Fourier components of generalized enthalpy density, and
\begin{eqnarray*}
\hat{\varepsilon}_{\vec{k}}=\sum_{\alpha}\hat{\varepsilon}_{\vec{k}}^{\alpha}
=\sum_{\alpha}\sum_{l=1}^{N_\alpha}{e}_{l}^{\alpha}e^{-ikr_l^{\alpha}}
\end{eqnarray*}
are the Fourier components of total energy density, where
\begin{eqnarray*}
{e}_{l}^{\alpha}=\frac{(\vec{p}^{\alpha}_l)^2}{2m_\alpha}
+\frac{1}{2}\sum_{\gamma}\sum_{j=1}^{N_\gamma}
V_{\alpha\gamma}(|\vec{r}_l^{\alpha}-\vec{r}_j^{\gamma}|).
\end{eqnarray*}
For these parameters of a reduced description  we can obtain a set of transport equations analogous to Eq. (\ref{math/IV.48}). However, now
the kinetic variable $\hat{N}_d(\vec{k},\vec{p},Z)$ is defines by the expression similar
to Eq. (\ref{math/IV.N})
\begin{eqnarray}
\hat{N}_d(\vec{k},\vec{p},Z)=(1-{\cal
P}^H)n^d_{\vec{k}}(\vec{p},Z)=(1-{\cal
P}^H)n^d_{\vec{k}}(\xi),
\end{eqnarray}
but the Mori projection operator ${\cal P}^H$ is constructed on the dynamic variables (\ref{math/IV.58})--(\ref{math/IV.59}). In this case,  the generalized hydrodynamic matrix can be written as follows (herewith, in the integral term
$\int d\xi T_{NN}(\vec{k},\xi,\xi') \langle \delta \hat{N}_d(\vec{k},\xi')\rangle^{t}$ in the equation for $\langle \delta \hat{N}_d(\vec{k},\xi)\rangle^{t}$ we make the approximation $T_{NN}(\vec{k},\xi,\xi')=T_{NN}(\vec{k},\xi,\xi)\delta (\xi-\xi')$ that corresponds to instantaneous transfer of impulse and charge, thus, $\int d\xi T_{NN}(\vec{k},\xi,\xi') \langle \delta \hat{N}_d(\vec{k},\xi')\rangle^{t}\approx T_{NN}(\vec{k},\xi,\xi)\langle \delta \hat{N}_d(\vec{k},\xi)\rangle^{t}$. Dynamic
variable $\langle \delta \hat{N}_d(\vec{k},\xi)\rangle^{t}$ for the grains  is a function of impulse and charge which change in the charging/discharging processes):
\begin{eqnarray}\label{math/IV.60}
T(k)=\left(\begin{array}{cccccc}
  \varphi_{nn}^{ee} & \varphi_{nn}^{ei} & \varphi_{nn}^{ea} & -i\Omega_{n\jmath}^{e} & \varphi_{nh}^{e} & \varphi_{nN}^{e} \\
  \varphi_{nn}^{ie} & \varphi_{nn}^{ii} & \varphi_{nn}^{ia} & -i\Omega_{n\jmath}^{i} & \varphi_{nh}^{i} & \varphi_{nN}^{i} \\
  \varphi_{nn}^{ae} & \varphi_{nn}^{ai} & \varphi_{nn}^{aa} & -i\Omega_{n\jmath}^{a} & \varphi_{nh}^{a} & \varphi_{nN}^{a} \\
 -i\Omega_{\jmath n}^{e} & -i\Omega_{\jmath n}^{i} & -i\Omega_{\jmath n}^{a} & \varphi_{\jmath\jmath} & -i\Omega_{\jmath h} & -i\Omega_{\jmath N} \\
 \varphi_{hn}^{e} & \varphi_{hn}^{i} & \varphi_{hn}^{a} & -i\Omega_{h\jmath} & \varphi_{hh} & \varphi_{hN} \\
\varphi_{Nn}^{e} & \varphi_{Nn}^{i} & \varphi_{Nn}^{a} & -i\Omega_{N\jmath} & \varphi_{Nh} & T_{NN} \\
\end{array}\right)_{(\vec{k},\vec{p},Z)}, \nonumber
\end{eqnarray}
where $T_{NN}=-i\Omega_{NN}+{\varphi_{NN}}$.

The calculation of collective modes reduces to finding the eigenvalues of this matrix.
Since it is difficult to obtain exact analytical expressions for eigenvalues of matrix (\ref{math/IV.60}) we use approximate calculations.
Conventionally, in the limit of small  $k$, eigenvalues can be found as a series over wave vector $z=z_0+z_1k+z_2k^2$. On the other hand, when certain cross-correlations are small
the perturbation theory for collective modes~\cite{23,25} can be developed.
Herewith, for the sake of simplification of calculations it is convenience
to pass from variables (\ref{math/IV.58})--(\ref{math/IV.59}) to the completely orthogonal set of variables:
\begin{align}\label{math/IV.61}
\hat{n}_{\vec{k}}^{(1)}&=\left(\hat{n}_{\vec{k}}^{e}-\frac{S_{ei}(k)}{S_{ii}(k)}\hat{n}_{\vec{k}}^{i}
-\frac{S_{ea}(k)}{S_{aa}(k)}\hat{n}_{\vec{k}}^{a}\right)/C_1(k),\\
\hat{n}_{\vec{k}}^{(2)}&=\left(\hat{n}_{\vec{k}}^{i}
-\frac{S_{ia}(k)}{S_{aa}(k)}\hat{n}_{\vec{k}}^{a}\right)/C_2(k),\\
\label{math/IV.61(3)}
\hat{n}_{\vec{k}}^{(3)}&=\hat{n}_{\vec{k}}^{a}/C_3(k),\\
\hat{\vec{J}}_{\vec{k}}\ &=\hat{\vec{\jmath}}_{\vec{k}}/
\langle\hat{\vec{\jmath}}_{\vec{k}}\hat{\vec{\jmath}}_{-\vec{k}}\rangle_0^{1/2},\\
\label{math/IV.61(5)}
\hat{H}_{\vec{k}}\
&=\hat{h}_{\vec{k}}/\langle\hat{h}_{\vec{k}}\hat{h}_{-\vec{k}}\rangle_0^{1/2},
\end{align}
where $S_{\alpha\gamma}(k)=\langle\hat{n}_{\vec{k}}^{\alpha}
\hat{n}_{-\vec{k}}^{\gamma}\rangle_0$ are the partial structure factors. Constants $C_\alpha(k)$ should provide the normalization $\langle\hat{n}_{\vec{k}}^{(\alpha)}
\hat{n}_{-\vec{k}}^{(\alpha)}\rangle_0=1$. The kinetic variable is orthogonal to this set.
Then, the generalized hydrodynamic matrix in new variables in determined by the relation
$\tilde{T}(k)=L^{-1}(k)T(k)L(k)$ and now it is symmetric ($L(k)$ denotes the matrix of a linear transformation):
\begin{eqnarray}\label{math/IV.61'}
{T}(k)=\left(\begin{array}{cccccc}
  {k^2 D_{11}} & k^2 D_{12} & k^2 D_{13} & -ik\omega_{nJ}^{(1)} & k^2\phi_{nH}^{(1)} & k^2\phi_{nN}^{(1)}(\xi) \\
  k^2 D_{21} & {k^2 D_{22}} & k^2 D_{23} & -ik\omega_{nJ}^{(2)} & k^2\phi_{nH}^{(2)} & k^2\phi_{nN}^{(2)}(\xi) \\
  k^2 D_{31} & k^2 D_{32} & {k^2 D_{33}} & -ik\omega_{nJ}^{(3)} & k^2\phi_{nH}^{(3)} & k^2\phi_{nN}^{(3)}(\xi) \\
 -ik\omega_{Jn}^{(1)} & -ik\omega_{Jn}^{(2)} & -ik\omega_{Jn}^{(3)} & {k^2D_l} & -ik\omega_{JH} & -ik\omega_{JN}(\xi) \\
 k^2\phi_{Hn}^{(1)} & k^2\phi_{Hn}^{(2)} & k^2\phi_{Hn}^{(3)} & -ik\omega_{HJ} & {k^2D_H} & k^2\phi_{HN}(\xi) \\
k^2\phi_{Nn}^{(1)}(\xi) & k^2\phi_{Nn}^{(2)}(\xi) & k^2\phi_{Nn}^{(3)}(\xi) & -ik\omega_{NJ}(\xi) & k^2\phi_{NH}(\xi) & -ik\omega_{NN}(\xi)+k^2D_{NN}(\xi)
 \\
\end{array}\right).\nonumber
\end{eqnarray}
Here, we extract the dependence of the frequency matrix and the matrix of memory functions on the wave vector in the hydrodynamic limit:
$i\tilde{\Omega}_{AA}(k)=ik\tilde{\omega}_{AA}$,
$\tilde{\varphi}_{AA}(k)=k^2\tilde{\phi}_{AA}$.
$D_{\alpha\gamma}$ are the corresponding diffusion coefficients,
$\phi_{Hn}^{(\alpha)}$ and $\phi_{nH}^{(\alpha)}$ describe thermodiffusion processes, $D_l$ is the longitudinal viscosity coefficient,
$D_H=\lambda/C_V$, where $C_V$ and $\lambda$ are the coefficients of heat capacity and heat conductivity, respectively.
$D_{NN}$ is the generalized diffusion coefficient for dust particles dependent on $\xi$.

\subsection*{Perturbation theory on correlations}

According to the perturbation theory for collective modes~\cite{23,25},
in order to calculate collective excitations in the zero approximation
we chose the matrix in the following form:
\begin{eqnarray}\label{math/IV.62}\small T_0(k)=
\left(\begin{array}{cccccc}
 {k^2 D_{11}} & 0 & 0 & 0 & 0 & 0 \\
  0 & {k^2 D_{22}} & 0 & 0 & 0 & 0 \\
  0 & 0 & {k^2 D_{33}} & -ik\omega_{nJ}^{(3)} & 0 & 0 \\
 0 & 0 & -ik\omega_{Jn}^{(3)} & {k^2D_l} & 0 & 0 \\
0 & 0 & 0 & 0 & {k^2D_H} & 0 \\
0 & 0 & 0 & 0 & 0 & -ik\omega_{NN}(\xi)+k^2D_{NN}(\xi) \\
\end{array}\right).
\end{eqnarray}
Such a choice is caused by the fact that dynamic variable $\hat{n}_3$
is related to the particles number density of neutral atoms (see (\ref{math/IV.61(3)})), concentration of which is
5--7 orders of magnitude larger than ion and electron concentration.
Respectively, the neutral component is the most energy-consuming and its contribution to the momentum density is the basic. That is why, the role of the neutral subsystem in the sound spreading is determinative. Visco-thermal correlations as well as correlations related to the dust component (due to small concentration) supposed to be small and can be taken into account as a perturbation.

Thus, in the zero approximation we obtain the spectrum of collective excitations and corresponding eigenvectors, namely:

-- two relaxation modes due to diffusion of plasma particles
\begin{align}
z_{D_1}^{(0)}&=-k^2D_{11}, & x_{D_1}&=(1,0,0,0,0,0),\label{math/IV.63}\\
z_{D_2}^{(0)}&=-k^2D_{22}, &
x_{D_2}&=(0,1,0,0,0,0);\label{math/IV.64}
\end{align}

-- complex conjugated sound modes
\begin{align}
z_{s_\pm}^{(0)}&=\pm \ ikc_0-k^2\Gamma_0,
& x_{s_\pm}&=\frac{1}{\sqrt{2}}(0,0,\pm
1,1,0,0),\label{math/IV.65}
\end{align}
where $c_0=\left(\omega_{Jn}^{(3)}\omega_{nJ}^{(3)}\right)^{1/2}$ is the isothermal sound velocity, $\Gamma_0=\frac{1}{2}(D_l+D_{33})$ is the sound damping coefficient in the zero approximation;

-- relaxation heat mode
\begin{align}
z_H^{(0)}&=-k^2D_H, & x_H&=(0,0,0,0,1,0);\label{math/IV.66}
\end{align}

-- relaxation mode related to diffusion of dust particles
\begin{align}
z_{N}^{(0)}&=ik\omega_{NN}(\xi)-k^2D_{NN}(\xi), &
x_{N}&=(0,0,0,0,0,1).\label{math/IV.67}
\end{align}

The cross-correlation define the perturbation matrix
\begin{eqnarray}
\label{math/IV.68} 
\delta T(k)=\left(\begin{array}{cccccc}
  0 & k^2 D_{12} & k^2 D_{13} & -ik\omega_{nJ}^{(1)} & k^2\phi_{nH}^{(1)} & k^2\phi_{nN}^{(1)}(\xi) \\
  k^2 D_{21} & 0 & k^2 D_{23} & -ik\omega_{nJ}^{(2)} & k^2\phi_{nH}^{(2)} & k^2\phi_{nN}^{(2)}(\xi) \\
  k^2 D_{31} & k^2 D_{32} & 0 & 0 & k^2\phi_{nH}^{(3)} & k^2\phi_{nN}^{(3)}(\xi) \\
 -ik\omega_{Jn}^{(1)} & -ik\omega_{Jn}^{(2)} & 0 & 0 & -ik\omega_{JH} & -ik\omega_{JN}(\xi) \\
 k^2\phi_{Hn}^{(1)} & k^2\phi_{Hn}^{(2)} & k^2\phi_{Hn}^{(3)} & -ik\omega_{HJ} & 0 & k^2\phi_{HN}(\xi) \\
k^2\phi_{Nn}^{(1)}(\xi) & k^2\phi_{Nn}^{(2)}(\xi) & k^2\phi_{Nn}^{(3)}(\xi) & -ik\omega_{NJ}(\xi) & k^2\phi_{NH}(\xi) & 0
 \\
\end{array}\right).\nonumber
\end{eqnarray}
Now we can calculate the corrections to collective modes caused by the weak cross-correlations. According to~\cite{23,25}, such corrections equal to zero in the first order in perturbation, and in the second order are determined by the formula
\begin{eqnarray}\label{math/IV.69}
\delta
z_\alpha^{(2)}=\sum_\beta\frac{\delta\bar{T}^*_{\alpha\beta}\delta\bar{T}_{\beta\alpha}}
{z_\alpha^{(0)}-z_\beta^{(0)}},
\end{eqnarray}
herewith, $\beta$ runs all possible values  but never equals $\alpha$. $\delta\bar{T}_{\alpha\beta}$ is the perturbation matrix in representation of eigenvectors of matrix $T_0(k)$.
$\delta\bar{T}_{\alpha\beta}=(x_\alpha\delta T
x_\beta)=\sum_{i,j}x_{i,\alpha}^*\delta T_{ij}x_{j,\beta}$, and
the conjugated matrix is determined by the relation
$\delta\bar{T}_{\alpha\beta}^*=\sum_{i,j}x_{i,\beta}^*\delta
T_{ij}x_{j,\alpha}$. When the necessary elements $\delta\bar{T}_{\alpha\beta}$ are calculated using Eqs. (\ref{math/IV.63})--(\ref{math/IV.67}) and taking into account the zero order results for collective modes Eqs.
(\ref{math/IV.63})--(\ref{math/IV.67}), we can calculate the desired corrections.
For diffusive modes they have the form
\begin{eqnarray}
\label{math/IV.70} \delta z_{D_1}^{(2)}=
-k^2\left(\frac{D_{12}D_{21}}{D_{11}-D_{22}}+\frac{\phi_{nH}^{(1)}\phi_{Hn}^{(1)}}
{D_{11}-D_H}+\omega_{Jn}^{(1)}\omega_{nJ}^{(1)}
\frac{\Gamma_0-D_{11}}{c_0^2}\right),
\end{eqnarray}
\begin{eqnarray}
\label{math/IV.71} \delta z_{D_2}^{(2)}=
-k^2\left(\frac{D_{21}D_{12}}{D_{22}-D_{11}}+\frac{\phi_{nH}^{(2)}\phi_{Hn}^{(2)}}
{D_{22}-D_H}+\omega_{Jn}^{(2)}\omega_{nJ}^{(2)}
\frac{\Gamma_0-D_{22}}{c_0^2}\right);
\end{eqnarray}
for sound modes we obtain
\begin{eqnarray}
\label{math/IV.72} \delta z_{s_\pm}^{(2)}&=&\pm \
ikc_0\left(\frac{\Delta}{2} \mp \frac{\omega_{JN}(\xi)\omega_{NJ}(\xi)}{2[\omega_{NN}(\xi)\mp c_0]} \right)\\
&-& k^2\frac{1}{2}\bigg\{-D_l\frac{\Delta}{2}\frac{1}{c_0^2}\bigg[
D_H\omega_{JH}\omega_{HJ}+D_{11}\omega_{Jn}^{(1)}\omega_{nJ}^{(1)}
+D_{22}\omega_{Jn}^{(2)}\omega_{nJ}^{(2)}\nonumber\\
&+& D_{33}\omega_{Jn}^{(3)}\omega_{nJ}^{(3)}
\left(1-\frac{\Delta}{2}\right)\bigg]
+\frac{\omega_{JN}(\xi)\omega_{NJ}(\xi)[\Gamma_0-D_{NN}(\xi)]}{2[\omega_{NN}(\xi) \mp c_0]^2} \bigg\} ,\nonumber
\end{eqnarray}
\begin{eqnarray*}
\Delta=\frac{\omega_{Jn}^{(1)}\omega_{nJ}^{(1)}+\omega_{Jn}^{(2)}\omega_{nJ}^{(2)}
+\omega_{JH}\omega_{HJ}} {c_0^2};
\end{eqnarray*}
the correction for heat mode
\begin{eqnarray}
\label{math/IV.73} \delta z_H^{(2)}=
-k^2\left(\frac{\phi_{Hn}^{(1)}\phi_{nH}^{(1)}}{D_H-D_{11}}+
\frac{\phi_{Hn}^{(2)}\phi_{nH}^{(2)}}{D_H-D_{22}}
+\omega_{HJ}\omega_{JH} \frac{\Gamma_0-D_H}{c_0^2}\right);
\end{eqnarray}
the correction for mode describing dust particles dynamics
\begin{eqnarray}
\label{math/IV.74}
\delta z_N^{(2)}&=&ik\omega_{NN}(\xi)\frac{\omega_{JN}(\xi)\omega_{NJ}(\xi)}{\omega_{NN}^2(\xi)-c_0^2}\\
&-&k^2\frac{\omega_{JN}(\xi)\omega_{NJ}(\xi)\left[\Gamma_0-D_{NN}(\xi)\right]
\left[\omega_{NN}^2(\xi)+c_0^2\right]}{\left[\omega_{NN}^2(\xi)-c_0^2\right]^2}.\nonumber
\end{eqnarray}
%

Now we are ready to write down the analytical expressions for collective excitations with corrections. Thus, diffusive modes have the following form
\begin{eqnarray}
\label{math/IV.70'}  z_{D_1}&=&
-k^2\left(D_{11}\left[1-\frac{\omega_{Jn}^{(1)}\omega_{nJ}^{(1)}}
{c_T^2}\right]\right.
+\left.
\frac{D_{12}D_{21}}{D_{11}-D_{22}}+\frac{\phi_{nH}^{(1)}\phi_{Hn}^{(1)}}
{D_{11}-D_H}+
\Gamma_0\frac{\omega_{Jn}^{(1)}\omega_{nJ}^{(1)}}{c_T^2}\right),
\end{eqnarray}
\begin{eqnarray}
\label{math/IV.71'} z_{D_2}&=&
-k^2\left(D_{22}\left[1-\frac{\omega_{Jn}^{(2)}\omega_{nJ}^{(2)}}
{c_T^2}\right]\right.
+\left.\frac{D_{21}D_{12}}{D_{22}-D_{11}}+\frac{\phi_{nH}^{(2)}\phi_{Hn}^{(2)}}
{D_{22}-D_H}+
\Gamma_0\frac{\omega_{Jn}^{(2)}\omega_{nJ}^{(2)}}{c_T^2}\right).
\end{eqnarray}
Sound excitations are as follows:
\begin{eqnarray}
\label{math/IV.72'} z_{s_\pm}&=&\pm \
ikc_0\left(1+\frac{\Delta}{2} \mp \frac{\omega_{JN}(\xi)\omega_{NJ}(\xi)}{2[\omega_{NN}(\xi)\mp c_0]} \right)\\
&-& k^2\frac{1}{2}\bigg\{D_l\left(1-\frac{\Delta}{2}\right)\frac{1}{c_0^2}\bigg[
D_H\omega_{JH}\omega_{HJ}+D_{11}\omega_{Jn}^{(1)}\omega_{nJ}^{(1)}
+D_{22}\omega_{Jn}^{(2)}\omega_{nJ}^{(2)}\nonumber\\
&+& D_{33}\omega_{Jn}^{(3)}\omega_{nJ}^{(3)}
\left(1-\frac{\Delta}{2}\right)\bigg]
+\frac{\omega_{JN}(\xi)\omega_{NJ}(\xi)[\Gamma_0-D_{NN}(\xi)]}{2[\omega_{NN}(\xi) \mp c_0]^2} \bigg\} .\nonumber
\end{eqnarray}
Heat mode has the form
\begin{eqnarray}
\label{math/IV.73'} z_H^{(2)}&=& -k^2\Biggl(D_H\left[1-
\frac{\omega_{HJ}\omega_{JH}}{c_T^2}\right]
+\frac{\phi_{Hn}^{(1)}\phi_{nH}^{(1)}}{D_H-D_{11}}+
\frac{\phi_{Hn}^{(2)}\phi_{nH}^{(2)}}{D_H-D_{22}}
+\Gamma_0\frac{\omega_{HJ}\omega_{JH}}{c_T^2}\Biggr).
\end{eqnarray}
For relaxation mode for dust particles we obtain
\begin{eqnarray}
\label{math/IV.74'}
z_N&=&ik\omega_{NN}(\xi)\left[1+\frac{\omega_{JN}(\xi)\omega_{NJ}(\xi)}{\omega_{NN}^2(\xi)-c_0^2}\right]\\
&-&k^2\left\{D_{NN}+\frac{\omega_{JN}(\xi)\omega_{NJ}(\xi)\left[\Gamma_0-D_{NN}(p)\right]
\left[\omega_{NN}^2(\xi)+c_0^2\right]}{\left[\omega_{NN}^2(\xi)-c_0^2\right]^2}
\right\}.\nonumber
\end{eqnarray}
%

Analyzing the obtained results we can see that taking into account cross-correlations
slightly modifies the collective modes of the system. Thus, contributions from dynamics of dust subsystem effect the sound velocity and coefficient of its attenuation in the system. Considering cross-correlations is manifested on the relaxation modes for dust particles.
Herewith, dynamics of grains does not effect diffusive and heat modes.
Considering cross-correlations we can take into account the dynamics of all components, in particular dust component, in the process of sound spreading and attenuation.
It is worth noting that sound modes in the second order in correlation are now not complex conjugated  (due to $[\omega_{NN}(\xi)\mp c_0]$ in Eq. (\ref{math/IV.71'})), and are characterized by the different velocity and sound damping.
Since $c_0$ is the isothermal sound velocity in the neutral component, then, $\omega_{NN}(\xi)$ can be defined as ``sound'' for dust subsystem dependent on  $\xi$. Then, the expression
$[\omega_{NN}(\xi)\mp c_0]$ in the denominators of terms for sound velocity and damping
strengthens or weakens acoustic excitations.
If we neglect the contribution of ``sound'' for dust ($[\omega_{NN}(\xi)\mp c_0]=0$)
we reproduce two conjugated collective modes.
It is important to note that sound modes and relaxation mode for grains depend on the value of charge and  impulse of dust grains, and, therefore change in the charging/discharging processes.

Neglecting the contribution from the dust subsystem we can reproduce the well-known expression for collective modes in multicomponent system within the hydrodynamic description~\cite{mryglod}. Within the description in terms of partial dynamics, besides modes mentioned above, we also obtain relaxation excitations of kinetic type related with interspecies interaction.


\section{Conclusions}

In summary, in the present paper we proposed another approach to the description
of such a complicated system as dusty plasma based on the Zubarev NSO method.
Since the subsystems of dusty plasma are in different states, such an approach allows us to take into account consistently kinetics of dust particles and hydrodynamics of electrons, ions and neutral atoms. Within the partial dynamics we constructed the nonequilibrium statistical operator of duty plasma and using it we obtained the generalized transport equations, which consistently take into account kinetic and hydrodynamic processes in the system. For the case of weakly nonequilibrium processes when the deviations of the nonequilibrium thermodynamic parameters from their equilibrium values are small, the set of transport equations is obtained. Analogous system of equations can be obtained for equilibrium time correlation functions built on the dynamic variables of a reduced description. In such an approximation this set of equations is closed.

Alternative to partial variables can serve variables of partial number densities, total momentum and total energy (enthalpy) densities. Such a set of variables can be used when investigate a spectrum of collective excitations of isothermal plasma in hydrodynamic limit. since investigation of collective modes in terms of partial variables is
more suitable beyond the of hydrodynamic region. Using the perturbation theory for collective modes in correlations we found a spectrum of collective excitations of dusty plasma. Herewith, in the second order in perturbation in diffusive and heat modes the terms connected with interaction of plasma particles and dust component do not appear.
Considering the dynamics of grains is manifested in the renormalization of sound velocity and dumping, and the latter are different for both modes.
Neglecting the correlations with dust subsystem we can reproduce the well-known expression for collective modes in multicomponent system~\cite{mryglod}.

\section*{APPENDIX A}
\renewcommand{\theequation}{A.\arabic{equation}}

First, we consider the case of partial dynamic variables.
In calculation of correlation function
\begin{eqnarray}
\Phi_{NN}(\vk,x,x')=\langle\hat{N}_\vk(\vec{p},Z)\hat{N}_{-\vk}(\vec{p}',Z')\rangle_0
=\langle(1-P_0)\hat{n}^d_\vk(\vec{p},Z)(1-P_0)\hat{n}^d_{-\vk}(\vec{p}',Z')\rangle_0
\end{eqnarray}
the static correlation function equals
\begin{eqnarray}
\langle\hat{n}^d_\vk(\vec{p},Z)\hat{n}^d_{-\vk}(\vec{p}',Z')\rangle_0
&=&n\delta(\vp-\vp')\delta(Z-Z')f_0(p)f_0(Z)\nonumber\\
&&+n_d^2f_0(p)f_0(p')f_0(Z)f_0(Z')h_2^d(k)
\end{eqnarray}
with equilibrium distributions in impulse and charge
\begin{eqnarray}
f_0(p)=\left(\frac{\beta}{2\pi m}\right)^{3/2}\exp\left(-\beta\frac{p^2}{2m}\right), \qquad
f_0(Z)=\left(\frac{\beta e^2}{\pi a_d}\right)^{1/2}\exp\left(-\beta\frac{Z_d^2e^2}{a_d}\right).
\end{eqnarray}

Taking into account new variable $\xi$ we can write down:
\begin{eqnarray}
\langle\hat{n}^d_\vk(\xi)\hat{n}^d_{-\vk}(\xi')\rangle_0
=n\delta(\xi-\xi')f_0(\xi)+n_d^2f_0(\xi)f_0(\xi')h_2^d(k)=\Phi_{nn}^d(\vk,\xi,\xi'),
\end{eqnarray}
where $h_2^d(k)$ is the correlation function for dust particles related with the direct correlation function
\begin{eqnarray}
h_2^d(k)=c_2^d(k)\left[1-n_dc_2^d(k)\right]^{-1}, \qquad [\Phi_{nn}^d(\vk,\xi,\xi')]^{-1}=\frac{\delta(\xi-\xi')}{n_df_0(\xi')}-c_2^d(k).
\end{eqnarray}
Let us calculate the action of the projection operator
\begin{eqnarray}
\hat{N}_\vk(\xi)=(1-P_0)\hat{n}^d_\vk(\xi)=\hat{n}^d_\vk(\xi)-P_0\hat{n}^d_\vk(\xi).
\end{eqnarray}
\begin{eqnarray}
P_0\hat{n}^d_\vk(\xi) & = & \sum_{\alpha,\gamma}\langle\hat{n}^d_\vk(\xi)\hat{n}_{-\vk}^\alpha\rangle_0
[\Phi_{nn}^{-1}(\vk)]^{\alpha\gamma}\hat{n}_\vk^\gamma
+\sum_{\gamma}\langle\hat{n}^d_\vk(\xi)\hat{\vec{\jmath}}_{-\vk}^\gamma\rangle_0
[\Phi_{\jmath\jmath}^{-1}(\vk)]^{\gamma\gamma}\hat{\vec{\jmath}}_\vk^\gamma \nonumber \\
&& + \sum_{\alpha,\gamma}\langle\hat{n}^d_\vk(\xi)\hat{h}_{-\vk}^\alpha\rangle_0
[\Phi_{hh}^{-1}(\vk)]^{\alpha\gamma}\hat{h}_\vk^\gamma \nonumber \\
& = & f_0(\xi)\sum_{\alpha,\gamma}\langle\hat{n}^d_\vk\hat{n}_{-\vk}^\alpha\rangle_0
[\Phi_{nn}^{-1}(\vk)]^{\alpha\gamma}\hat{n}_\vk^\gamma
+ f_0(\xi)\sum_{\alpha,\gamma}\langle\hat{n}^d_\vk\hat{h}_{-\vk}^\alpha\rangle_0
[\Phi_{hh}^{-1}(\vk)]^{\alpha\gamma}\hat{h}_\vk^\gamma
\end{eqnarray}
with partial structure factors
\begin{eqnarray}
\Phi^{d\gamma}(k)=\langle\hat{n}^d_\vk  \hat{n}_{-\vk}^\gamma\rangle_0, \qquad
\Phi^{d\gamma}_{nh}(k)=\langle\hat{n}^d_\vk  \hat{h}_{-\vk}^\gamma\rangle_0.
\end{eqnarray}
Introducing the normalized static correlation functions
\begin{eqnarray}
\label{eq:norm}
\bar{\Phi}^{d\gamma}_{nn}(\vk)=\sum_\alpha\Phi_{nn}^{d\alpha}[\Phi_{nn}^{-1}(\vk)]^{\alpha\gamma}, \qquad
\bar{\Phi}^{d\gamma}_{nh}(\vk)=\sum_\alpha\Phi_{nh}^{d\alpha}[\Phi_{hh}^{-1}(\vk)]^{\alpha\gamma},
\end{eqnarray}
we obtain
\begin{eqnarray}
P_0\hat{n}^d_\vk(\xi)&=&f_0(\xi)\sum_\gamma \left\{\bar{\Phi}_{nn}^{d\gamma}(\vk)\hat{n}^{\gamma}_\vk+
\bar{\Phi}_{nh}^{d\gamma}(\vk)\hat{h}^{\gamma}_\vk\right\},\nonumber \\
P_0\hat{n}^d_{-\vk}(\xi')&=&f_0(\xi')\sum_\gamma \left\{\bar{\Phi}_{nn}^{d\gamma}(\vk)\hat{n}^{\gamma}_{-\vk}+
\bar{\Phi}_{nh}^{d\gamma}(\vk)\hat{h}^{\gamma}_{-\vk}\right\}.
\end{eqnarray}
Further, we calculate contributions into the correlation function
\begin{eqnarray}
\Phi_{NN}(\vk,\xi,\xi')=\langle\hat{N}_\vk(\xi)\hat{N}_{-\vk}(\xi')\rangle_0&=&
\langle\hat{n}^d_\vk(\xi)\hat{n}^d_{-\vk}(\xi')\rangle_0
-\langle\hat{n}^d_\vk(\xi)P_0\hat{n}^d_{-\vk}(\xi')\rangle_0\nonumber\\
&&-\langle P_0\hat{n}^d_\vk(\xi)\hat{n}^d_{-\vk}(\xi')\rangle_0
+\langle P_0\hat{n}^d_\vk(\xi)P_0\hat{n}^d_{-\vk}(\xi')\rangle_0.
\end{eqnarray}
%
%
Taking into account Eq. (\ref{eq:norm}) it easily to show that
%
%
\begin{eqnarray}
\langle\hat{n}^d_\vk(\xi)P_0\hat{n}^d_{-\vk}(\xi')\rangle_0=
\langle P_0\hat{n}^d_\vk(\xi)\hat{n}^d_{-\vk}(\xi')\rangle_0=\langle P_0\hat{n}^d_\vk(\xi)P_0\hat{n}^d_{-\vk}(\xi')\rangle_0=f_0(\xi)f_0(\xi')G(k),
\end{eqnarray}
where
\begin{eqnarray}
G(k)=\sum_{\gamma}\Big\{\bar{\Phi}_{nn}^{d\gamma}(\vk)\Phi_{nn}^{\gamma d}(\vk)
+\bar{\Phi}_{nh}^{d\gamma}(\vk)\Phi_{nh}^{\gamma d}(\vk)\Big\}.
\end{eqnarray}

Finally, we obtain
\begin{eqnarray}
\Phi_{NN}(\vk,\xi,\xi')&=&
n_d\delta(\xi-\xi')f_0(\xi')+f_0(\xi)f_0(\xi')[n_d^2h_2^d(k)-G(k)]\nonumber\\
&=&
n_d\delta(\xi-\xi')f_0(\xi')+f_0(\xi)f_0(\xi')\bar{G}(k),
\end{eqnarray}
where
\begin{eqnarray}
\bar{G}(k)=n_d^2h_2^d(k)-G(k).
\end{eqnarray}
Function inverse to $\Phi_{NN}(\vk,\xi,\xi')$ we find from the condition
\begin{eqnarray}
\int d\xi''\Phi_{NN}^{-1}(\vk,\xi,\xi'')\Phi_{NN}(\vk,\xi'',\xi')=\delta(\xi-\xi').
\end{eqnarray}
The final result is
\begin{eqnarray}
\Phi_{NN}^{-1}(\vk,\xi,\xi')=\frac{\delta(\xi-\xi')}{n_df_0(\xi')}-B(k), \qquad B(k)=\frac{n_d^2\bar{G}(k)}{n_d\left[n_d+\bar{G}(k)\right]}.
\end{eqnarray}

In the case of conservative variables, for $\Phi_{NN}^{-1}(\vk,\xi,\xi')$ we obtain similar expression, the only difference is $\bar{G}(k)$ replaced with $\bar{G}_\mathrm{c}(k)$, where
\begin{eqnarray}
\bar{G}_\mathrm{c}(k)=n_d^2h_2^d(k)-G_\mathrm{c}(k),
\qquad
G_\mathrm{c}(k)=\sum_{\gamma}\Big\{\left(\Phi_{nn}^{\gamma d}(\vk)\right)^2
+\left(\Phi_{nH}^{\gamma}(\vk)\right)^2\Big\}.
\end{eqnarray}
%

\section*{APPENDIX B}
\renewcommand{\theequation}{B.\arabic{equation}}

Here, we calculate the elements of the generalized hydrodynamic matrix (\ref{math/IV.61'}).  Let start with frequency matrix $i\Omega_{AB}(k)=ik\,\omega_{AB}(k)$, which in new completely orthogonal variables (\ref{math/IV.61})--(\ref{math/IV.61(5)}) is symmetric and whose elements are defined by the relation
\begin{eqnarray*}
i\Omega_{AB}(k)=\langle\dot{\hat{A}}_{\vk}\hat{B}_{-\vk}\rangle_0\,.
\end{eqnarray*}
For instance, let us consider
\begin{eqnarray}\label{B.1}
i\Omega_{nJ}^{(3)}(k)=i\Omega_{Jn}^{(3)}(k)=\langle\dot{\hat{n}}_{\vk}^{(3)}\hat{\vec{J}}_{-\vk}\rangle_0
=\frac{\langle\dot{\hat{n}}_{\vk}^{a}
\hat{\vec{\jmath}}_{-\vk}\rangle_0}
{\langle\hat{n}_{\vk}^{a}\hat{n}_{-\vk}^{a}\rangle_0^{1/2}
\langle\hat{\vec{\jmath}}_{\vk}\hat{\vec{\jmath}}_{-\vk}\rangle_0^{1/2}}
=\frac{ik}{m_a}\frac{\langle\hat{\vec{\jmath}}_{\vk}^{\ a}
\hat{\vec{\jmath}}_{-\vk}^{\ a}\rangle_0}
{\langle\hat{n}_{\vk}^{a}\hat{n}_{-\vk}^{a}\rangle_0^{1/2}
\langle\hat{\vec{\jmath}}_{\vk}\hat{\vec{\jmath}}_{-\vk}\rangle_0^{1/2}}.
\end{eqnarray}
The static correlation function can be easily calculated
\begin{eqnarray}
\langle\hat{\vec{\jmath}}_{\vk}\hat{\vec{\jmath}}_{-\vk}\rangle_0
=\sum_l \langle\hat{\vec{\jmath}}_{\vk}^{\ l}\hat{\vec{\jmath}}_{-\vk}^{\ l}\rangle_0
=\frac{1}{\beta}\sum_l N_l m_l
=\frac{1}{\beta}\left(N_a m_a + N_i m_i + N_e m_e\right),
\end{eqnarray}
where the index $l=a,i,e$.
Taking into account that in our case
\begin{eqnarray}
m_a \simeq m_i \gg m_e, \qquad N_a \gg N_i \simeq N_e
\end{eqnarray}
we obtain an approximate result
\begin{eqnarray}
\langle\hat{\vec{\jmath}}_{\vk}\hat{\vec{\jmath}}_{-\vk}\rangle_0
\simeq \frac{N_a m_a}{\beta}\, .
\end{eqnarray}


Thus, we obtain for the following elements:
\begin{align}
\label{B.2}
i\Omega_{nJ}^{(3)}(k)&=i\Omega_{Jn}^{(3)}(k)
=-\frac{ik}{m_\alpha}\frac{\langle\hat{\vec{\jmath}}_{\vk}^{\ a}
\hat{\vec{\jmath}}_{-\vk}^{\ a}\rangle_0^{1/2}}
{\langle\hat{n}_{\vk}^{a}\hat{n}_{-\vk}^{a}\rangle_0^{1/2}}
=-ik\left(\frac{N_a}{\beta m_a S_{aa}}\right)^{1/2},\\
%
\label{B.3}
i\Omega_{nJ}^{(2)}(\vk)&=i\Omega_{Jn}^{(2)}(\vk)
=-ik\left(\frac{N_a}{\beta m_a C_{2}}\right)^{1/2}\left(\frac{N_i}{N_a}-\frac{f_{ia}}{f_{aa}}\right),\\
%
\label{B.4}
i\Omega_{nJ}^{(1)}(\vk)&=i\Omega_{Jn}^{(1)}(\vk)
=-ik\left(\frac{N_a}{\beta m_a C_{1}}\right)^{1/2}\left(\frac{N_e}{N_a}-\alpha\frac{N_i}{N_a}-\delta\right),\\
%
\label{B.5}
i\Omega_{HJ}(\vk)&=i\Omega_{JH}(\vk)
=-ik\frac{\alpha_P}{\varkappa_T}\left(\frac{TV}{n\bar{m}C_V}\right)^{1/2}.
\end{align}

Here, we use the following notations:
\begin{align*}
C_1&=\frac{1}{S_{aa}}\left[S_{aa}S_{ee}- S_{ea}^2
- \frac{(S_{aa}S_{ei}-S_{ia}S_{ea})^2}{S_{aa}S_{ii}-S_{ia}^2}\right]\, , \\
C_2&=S_{ii}-\frac{S_{ia}^2}{S_{aa}}\, ,
\qquad
C_3=S_{aa}\, , \\
\delta_1&=\frac{S_{ia}}{S_{aa}}\, , 
\qquad
\delta_2=\frac{S_{aa}S_{ei}-S_{ia}S_{ea}}{S_{aa}S_{ii}-S_{ia}^2}\, , 
\qquad
\delta_3=\frac{S_{ea}}{S_{aa}}-\delta_1\delta_2\,.
\end{align*}

The elements of the matrix of transport kernels in new variables are defined by the relation
$k^2\phi_{AB}(\vk;t,t')=\langle I_A(\vk)T_0(t,t')I_B(\vk)\rangle_0$ and in our case should be calculated in the Markovian approximation:
\begin{align}
\label{B.6}
\phi_{33}&=D_{aa}S_{aa}^{-1},\\
%
\label{B.7}
\phi_{22}&=C_{2}^{-1}\left[D_{ii}-2\delta_1D_{ia}
+\delta_1^2D_{aa}\right],\\
%
\label{B.8}
\phi_{23}&=\left(C_{2}C_{3}\right)^{-1/2}\left[D_{ia}-\delta_1D_{aa}\right],\\
%
\label{B.9}
\phi_{13}&=\left(C_{1}C_{3}\right)^{-1/2}\left[D_{ea}-\delta_2 D_{ia}-\delta_3 D_{aa}\right],\\
%
\label{B.10}
\phi_{12}&=\left(C_{1}C_{2}\right)^{-1/2}
\left[D_{ei}-\delta_1 D_{ea}-\delta_2 D_{ii}+\left(\delta_2-\delta_1 \right)D_{ia}+\delta_1\delta_3 D_{aa}\right],\\
%
\label{B.11}
\phi_{11}&=\left(C_{1}\right)^{-1}
\left[D_{ee}+\delta_2^2 D_{ii}+\delta_3^2 D_{aa} - 2\delta_2 D_{ei}- 2\delta_3 D_{ea}
+2\delta_2\delta_3 D_{ia}\right],
\end{align}
where $D_{ll'}$ are the diffusion coefficients.

The transport kernels
$
\label{B.12}
\phi_{nH}^{(\alpha)}=\phi_{Hn}^{(\alpha)}=\langle I_n^{(\alpha)}T_0(t,t')I_{H}\rangle_0
$
have the form
\begin{align}
\label{B.13}
\phi_{nH}^{(3)}&=\phi_{Hn}^{(3)}=K_a\left(\frac{k_B}{C_V S_{aa}}\right)^{1/2},\\
%
\label{B.14}
\phi_{nH}^{(2)}&=\phi_{Hn}^{(2)}=\left(K_i-\delta_1K_a\right)\left(\frac{k_B}{C_V C_2}\right)^{1/2},\\
%
\label{B.15}
\phi_{nH}^{(1)}&=\phi_{Hn}^{(1)}=\left(K_e-\delta_2 K_i-\delta_3 K_a\right)\left(\frac{k_B}{C_V C_1}\right)^{1/2}
\end{align}
and are expressed via thermodiffusion coefficients $K_l$. The transport kernels
\begin{align}\label{B.16}
\phi_{JJ}&=\frac{\eta_L}{\bar{m} n}\,,\\
%
\phi_{HH}&=\frac{\lambda}{T c_v}\,
\end{align}
define coefficients of longitudinal viscosity and heat conductivity.

Further, we calculate the correlations between plasma particles and the grains subsystem.
For coefficients  $\phi_{Nn}^{(\alpha)}$ we obtain:
\begin{eqnarray}
\label{B.17}
\phi_{Nn}^{(\alpha)}(\xi)=\phi_{nn}^{(\alpha)}(\xi)
-f_0(\xi)\sum_{\beta}\Phi_{nn}^{d\beta}\phi_{\beta\alpha}
-f_0(\xi)\Phi_{nH}^{d}\phi_{Hn}^{(\alpha)},
\end{eqnarray}
where $\phi_{\beta\alpha}$ are defined by the relations  (\ref{B.6})--(\ref{B.11}), а $\phi_{Hn}^{(\alpha)}$ -- by Eqs. (\ref{B.13})--(\ref{B.15}). $\phi_{nn}^{(\alpha)}(\xi)$ is the diffusion coefficient (in momentum space) constructed  similarly to Eqs. (\ref{B.6})--(\ref{B.11}).

In analogous way
\begin{eqnarray}
\label{B.18}
\phi_{nN}^{(\alpha)}(\xi)=\int \Big\{ \phi_{nn}^{(\alpha)}(\xi')
-f_0(\xi')\sum_{\beta}\Phi_{nn}^{d\beta}\phi_{\beta\alpha}
-f_0(\xi')\Phi_{nH}^{d}\phi_{Hn}^{(\alpha)}\Big\}\Phi_{NN}^{-1}(\xi',\xi)d\xi'.
\end{eqnarray}

The generalized diffusion coefficient for dust particles has the following form
\begin{eqnarray}
\label{B.19}
\phi_{NN}(\xi,\xi')&=&\int \Big\{ \phi_{nn}(\xi,\xi'')
-f_0(\xi'')\sum_{\beta}\Phi_{nn}^{d\beta}\phi_{\beta}(\xi)
-f_0(\xi)\sum_{\beta}\Phi_{nn}^{d\beta}\phi_{\beta}(\xi'')\\
&&-f_0(\xi'')\Phi_{nH}^{d}\phi_{nH}(\xi)
-f_0(\xi)\Phi_{nH}^{d}\phi_{nH}(\xi'')
+f_0(\xi)f_0(\xi'')\sum_{\alpha,\beta}\Phi_{nn}^{d\alpha}\Phi_{nn}^{d\beta}
\phi_{\alpha\beta}\nonumber\\
&&+2f_0(\xi)f_0(\xi'')\sum_{\alpha}\Phi_{nn}^{d\alpha}\Phi_{nH}^{d}
\phi_{nH}^{\alpha}
+f_0(\xi)f_0(\xi'')(\Phi_{nH}^{d})^2
\phi_{HH}
\Big\}\Phi_{NN}^{-1}(\xi',\xi)d\xi',\nonumber
\end{eqnarray}
where $\phi_{nn}(\xi,\xi'')$ is a ``pure'' diffusion coefficient in momentum space for dust grains.

For element $i\Omega_{NN}=ik\omega_{NN}$ we can obtain
\begin{eqnarray}
\label{B.20}
i\Omega_{NN}(xi,xi')&=&-\frac{i\vec{k}\cdot\vec{p}}{m_d}
\Bigg\{
\frac{f_0(\xi)\delta(\xi-\xi')}{f_0(\xi')}\\
&&+n_df_0(\xi)\left[ h_2^d(k)-B_(k)\right]
-n_d^2f_0(\xi)B(k)h_2^d(k)
\Bigg\}.\nonumber
\end{eqnarray}

The last functions to be calculated are
\begin{eqnarray}\label{B.21}
i\Omega_{NJ}(xi)=-f_0(\xi)\left[
\sum_\alpha\Phi_{nn}^{d\alpha}i\Omega_{nJ}^{\alpha}
+\Phi_{nH}^{d}i\Omega_{HJ}\right],\nonumber
\end{eqnarray}
\begin{eqnarray}\label{B.22}
i\Omega_{JN}(xi)=-\left[
\sum_\alpha\Phi_{nn}^{d\alpha}i\Omega_{nJ}^{\alpha}
+\Phi_{nH}^{d}i\Omega_{HJ}\right]\frac{1-n_d B(k)}{n_d},\nonumber
\end{eqnarray}
where elements $i\Omega_{nJ}$ and $i\Omega_{HJ}$ are defined by the relations Eqs. (\ref{B.2})--(\ref{B.5}).


\begin{thebibliography}{99}

\bibitem{1}  V.N. Tsytovich, J. Winter, Phys.-Usp., 1998, \textbf{41}, 815.

\bibitem{2}  V.N. Tsytovich, Phys.-Usp., 1997, \textbf{40}, 53.

\bibitem{2a} Clark R.A., Sheldon R.B.
Dusty plasma based fission fragment nuclear reactor. In: 41 st AIAA/ASMA/SAE/ASEE, Joint
Propulsion Conf., Exhibit, July 10-15, 2005, Tucson, AZ, p.1-7.

\bibitem{3} V.E. Fortov, A.G. Khrapak, S.A. Khrapak, V.I. Molotkov, O.F. Petrov,  Phys.-Usp., 2004, \textbf{47}, 447.

\bibitem{4}  Fortov V.E.,  Ivlev A.V.,  Khrapak S.A.,  Khrapak A.G.,  Morfill G.E., Phys. Rep., 2005, \textbf{421}, 1.

\bibitem{5}  Tsytovich V.N.,  Phys.-Usp., 2007, \textbf{50}, 409.

\bibitem{Bonitz1} Introduction to Complex Plasmas, Tutorial lectures
M. Bonitz, N. Horing, and P. Ludwig (eds.)
Springer Series "Atomic, Optical and Plasma Physics", vol. 59, Springer, Berlin, 2010.

\bibitem{Bonitz2} Melzer A.,  Schella A.,  Miksch T.,  Schablinski J.,  Block D.,  Piel A.,  Thomsen H.,  Ka"hlert H., and M. Bonitz,
Contrib. Plasma Phys.,2012, vol. 52 , No. 10, p.795-803.

\bibitem{5a} Tsytovich V.N., JETP Lett., 2003, \textbf{78}, 763

\bibitem{6}  Bystrenko~T., Zagorodny~A.G., Ukr. J.~Phys., 2002, \textbf{47}, 431.

\bibitem{7}  Bashkirov A.G.,
Phys. Rev. E, 2004, vol. 69, No 4, 046410.

\bibitem{44}  Sitenko~A.G., Zagorodny~A.G., and Tsytovich~V.N.,
             In: Int. Conf. on Plasma Physics ICPP 1994, (ed.)
             P.H.Sakanaka and M.Tendler, AIP Conf. Proc., 1995, 311.

\bibitem{55}  Sitenko~A.G., Zagorodny~A.G., Chutov~Yu.I., Schram~P.P.J.M.
             and Tsytovich~V.N., Plasma Phys. Controlled Fusion, 1996, \textbf{38}, A105.

\bibitem{66}  Schram~A.A., Sitenko~A.G., Trigger~S.A.,
             Zagorodny~A.G., Phys. Rev.~E, 2000, \textbf{63}, 016403.

\bibitem{8}  Tsytovich~V.N. and Havnes~O., Comments Plasma Phys.
             Control. Fusion, 1993, \textbf{15}, 267.

\bibitem{9}  Wang~X. and Bhattacharje~A., Phys. Plasmas, 1996, \textbf{3}, 1189.

\bibitem{10} Tsytovich~V.N., de~Angelis~U., Bingham~R.,
             Resendes~D., Phys. Plasmas, 1997, \textbf{4}, 3882.

\bibitem{11} Ignatov~A.M., J.~Phys.~IV, 1997, \textbf{C4}, 215.

\bibitem{12} Trigger~S.A. and Schram~P.P.J.M., J.~Phys.~D, 1999, \textbf{32}, 234.

\bibitem{13} Tsytovich~V.N., de~Angelis~U., Phys. Plasmas, 1999, \textbf{6}, 1093.

\bibitem{33} Tsytovich~V.N., de~Angelis~U., Phys. Plasmas, 2000, \textbf{7}, 554.

\bibitem{43} Tsytovich~V.N., de~Angelis~U., Phys. Plasmas, 2001, \textbf{8}, 1141.

\bibitem{53} Tsytovich~V.N., de~Angelis~U., Phys. Plasmas, 2002, \textbf{9}, 2497.

\bibitem{63} Tsytovich~V.N., de~Angelis~U., Phys. Plasmas, 2004, \textbf{11}, 496.

\bibitem{73} Tsytovich~V.N., de~Angelis~U., Phys. Plasmas, 2005, \textbf{12}, 112311.

\bibitem{14} Zagorodny~A.G., Ukr. J.~Phys., 2005, \textbf{50}, 205.

\bibitem{15} Bystrenko~O., Zagorodny~A., Phys. Rev.~E, 2003, \textbf{67}, 066403.

\bibitem{16} Bystrenko~O., Bystrenko~T., Zagorodny~A.G., Phys. Lett.~A,
             2004, \textbf{329}, 83.

\bibitem{17} Bystrenko~O., Bystrenko~T., Zagorodny~A.G.,
             Ukr. J.~Phys., 2005, \textbf{50}, 557.

\bibitem{18} Zagorodny~A., Mal'nev~V., Rumyantsev~S.,
             Ukr. J.~Phys., 2005, \textbf{50}, 448.

\bibitem{sem} Semenov I.L., Zagorodny~A.G., Krivtsun I.V., Phys. Plasmas, 2013, \textbf{20}, No 1, 013701.

\bibitem{Trib1} Amour R., and Tribeche M. Variable charge dust acoustic solitary waves
in a dusty plasma with a $q$-nonextensive electron velocity distribution.// Phys. Plasmas, 2010, vol. 17, No 6, p.063702(1-7)

\bibitem{Trib2}Tribeche M., Shukla P.K. Charging of a dust particle in a plasma with a non extensive electron distribution function.
// Phys. Plasmas, 2011, vol. 18, No 10, p.103702(1-4)

\bibitem{Du} Jindyu G., Du J. Dust charging processes in the nonequilibrium dusty plasma with nonextensive power-law distribution.
arXiv:1202.0636, 2012 - arxiv.org, 16p.

\bibitem{Fort1} Vaulina O.S., Gavrikov A.V., Shakhova I.A., Petrov O.F., Vorona N.A., Fortov V.E. Study of Relation between Transport Coefficients in Dusty Plasma Systems. 32nd EPS Conf.Plasma Phys. Tarragona, 27 June-1July, 2005 ECA, vol. 29C, P-2.136, 4p.

\bibitem{19} Zubarev D.N., Morozov V.G., Ropke G., Statistical Mechanics
             of Nonequilibrium Processes, Vol.~1. Akad. Verlag, Berlin, 1996.

\bibitem{19'}Zubarev D.N., Morozov V.G., Ropke G., Statistical Mechanics
             of Nonequilibrium Processes, Vol.~2. Akad. Verlag, Berlin, 1997.

\bibitem{23} Bryk T.M., Mryglod I.M., Condens. Matter Phys., 2008, \textbf{11}, 139.

\bibitem{25} Mryglod I.M., Kuporov V.M., Ukr. J.~Phys., 2010, \textbf{55}, 1172.

\bibitem{22} Tokarchuk M.V., Markiv B.B., Collected Physical Papers, Shevchenko Scientific Society, 2008, \textbf{7}, 100 (in Ukrainian).

\bibitem{CMP_2010} Markiv B.B., Omelyan I.P., Tokarchuk M.V.,
Condens. Matter Phys., 2010, \textbf{13}, 23005.

\bibitem{kost} Kostrobii P.P., Tokarchuk M.V., Markovych B.M., Ignatyuk V.V., Gnativ B.V.,
Reaction-diffusion processes in the ``metal-gas'' systems. Publishing house of National University ``Lviv Polytechnic'', Lviv, 2009 (in Ukrainian).

\bibitem{mryglod} Mryglod I.M., Condens. Matter Phys., 1998, \textbf{1}, 753.

\end{thebibliography}
\end{document}